\documentclass{aa}  

\usepackage{graphicx}
\usepackage{dsfont}
\usepackage{txfonts}
\usepackage[normalem]{ulem}
\usepackage{color}
\usepackage{bm}
\usepackage{booktabs} % added by WHB: makes nicer looking tables
\usepackage[dvipsnames]{xcolor}
\begin{document}

   \title{{\it Kepler} Observations of the Asteroseismic Binary HD 176465}

   \author{T.~R.~White\inst{1,2,3}
          \and
          O.~Benomar\inst{4,5}
          \and
          V.~Silva~Aguirre\inst{1}
          \and
          W.~H.~Ball\inst{2,3}
          \and
          T.~R.~Bedding\inst{6,1}
          \and
          W.~J.~Chaplin\inst{7,1}
          \and
          J.~Christensen-Dalsgaard\inst{1}
          \and
          R.~A.~Garcia\inst{8}
          \and
          L.~Gizon\inst{3,2,5}
          \and
          D.~Stello\inst{6,1}
          \and
          S.~Aigrain\inst{9}
          \and
          H.~M.~Antia\inst{10}
          \and
          T.~Appourchaux\inst{11}
          \and
          M.~Bazot\inst{5}
          \and
          T.~L.~Campante\inst{7,1}
          \and
          O.~L.~Creevey\inst{12}
          \and
          G.~R.~Davies\inst{7,1,8}
          \and
          Y.~P.~Elsworth\inst{7,1}
          \and
          P.~Gaulme\inst{13,14}
          \and
          R.~Handberg\inst{1}
          \and
          S.~Hekker\inst{3,1}
          \and
          G.~Houdek\inst{1}
          \and
          R.~Howe\inst{7}
          \and
          D.~Huber\inst{6,15,1}
          \and
          C.~Karoff\inst{1,16}
          \and
          J.~P.~Marques\inst{17}
          \and
          S.~Mathur\inst{18}
          \and
          A.~McQuillan\inst{19}
          \and
          T.~S.~Metcalfe\inst{18}
          \and
          B.~Mosser\inst{20}
          \and
          M.~B.~Nielsen\inst{2,3}
          \and
          C.~R\'egulo\inst{21,22}
          \and
          D.~Salabert\inst{8}
          \and
          T.~Stahn\inst{2,3}
          }

   \institute{Stellar Astrophysics Centre, Department of Physics and Astronomy, Aarhus University, Ny Munkegade 120, DK-8000 Aarhus C, Denmark;
   \email{tim@phys.au.dk}
   \and
        Institut f\"{u}r Astrophysik, Georg-August-Universit\"{a}t G\"{o}ttingen, Friedrich-Hund-Platz 1, 37077 G\"{o}ttingen, Germany                
   \and
        Max-Planck-Institut f\"ur Sonnensystemforschung, Justus-von-Liebig-Weg 3, 37077 G\"ottingen, Germany   
   \and
        Department of Astronomy, The University of Tokyo, School of Science, Tokyo 113-0033, Japan
    \and
    	Center for Space Science, NYUAD Institute, New York University Abu Dhabi, PO Box 129188, Abu Dhabi, UAE
    \and
        Sydney Institute for Astronomy (SIfA), School of Physics, University of Sydney, NSW 2006, Australia
    \and
        School of Physics and Astronomy, University of Birmingham, Birmingham B15 2TT, UK
    \and
        Laboratoire AIM, CEA/DRF – CNRS – Univ. Paris Diderot – IRFU/SAp, Centre de Saclay, 91191 Gif-sur-Yvette Cedex, France
    \and
        Oxford Astrophysics, University of Oxford, Keble Rd, Oxford OX1 3RH, UK
    \and
        Tata Institute of Fundamental Research, Homi Bhabha Road, Mumbai 400005, India
    \and
        Univ. Paris-Sud, Institut d'Astrophysique Spatiale, UMR 8617, CNRS, B\^{a}timent 121, 91405 Orsay Cedex, France
    \and
        Laboratoire Lagrange, Universit\'e C\^ote d'Azur, Observatoire de la C\^ote d'Azur, CNRS, Blvd de l'Observatoire, CS 34229, 06304 Nice cedex 4, France
    \and
        Department of Astronomy, New Mexico State University, P.O. Box 30001, MSC 4500, Las Cruces, NM 88003-8001, USA
    \and
        Apache Point Observatory, 2001 Apache Point Road, P.O. Box 59, Sunspot, NM 88349, USA
    \and
        SETI Institute, 189 Bernardo Avenue, Suite 100, Mountain View, CA 94043, USA
    \and
        Department of Geoscience, Aarhus University, H{\o}egh-Guldbergs Gade 2, 8000, Aarhus C, Denmark
    \and
        LESIA, CNRS, Universit\'e Pierre et Marie Curie, Universit\'e Denis Diderot, Observatoire de Paris, 92195 Meudon Cedex, France 
    \and
        Space Science Institute, 4750 Walnut St. Suite 205, Boulder CO 80301, USA
    \and
        School of Physics and Astronomy, Raymond and Beverly Sackler, Faculty of Exact Sciences, Tel Aviv University, 69978, Tel Aviv, Israel
    \and
        LESIA, Observatoire de Paris, PSL Research University, CNRS, Sorbonne Universit\'es, UPMC Univ. Paris 06, Univ. Paris Diderot, Sorbonne Paris Cit\'e
    \and
        Instituto de Astrof\'isica de Canarias, 38205 La Laguna, Tenerife, Spain
    \and
        Universidad de La Laguna, Dpto. de Astrof\'isica, 38206 La Laguna, Tenerife, Spain
             }

   \date{Received Month Day, Year; accepted Month Day, Year}

\abstract{Binary star systems are important for understanding stellar structure and evolution, and are especially useful when oscillations can be detected and analysed with asteroseismology. However, only four systems are known in which solar-like oscillations are detected in both components. Here, we analyse the fifth such system, HD 176465, which was observed by {\it Kepler}. We carefully analysed the system's power spectrum to measure individual mode frequencies, adapting our methods where necessary to accommodate the fact that both stars oscillate in a similar frequency range. We also modelled the two stars independently by fitting stellar models to the frequencies and complementary parameters. We are able to cleanly separate the oscillation modes in both systems. The stellar models produce compatible ages and initial compositions for the stars, as is expected from their common and contemporaneous origin. Combining the individual ages, the system is about $3.0\pm0.5\,\mathrm{Gyr}$ old. The two components of HD 176465 are young physically-similar oscillating solar analogues, the first such system to be found, and provide important constraints for stellar evolution and asteroseismology.}%{}{}{}{} 
% 5 {} token are mandatory
 
%   \abstract
%   {}


   \maketitle
%
%________________________________________________________________

\section{Introduction}
Observations of binary stars have been critical in advancing our knowledge of stellar structure and evolution. The shared formation history of stars in these systems results in their having the same initial chemical composition and age. Additionally, the dynamics of the system can reveal the component masses, and their radii if they eclipse. 

Studies of binary stars can be further enhanced by asteroseismic measurements. Asteroseismology of solar-like stars is rapidly proving its effectiveness at characterising stars, driven by the successes of the CoRoT and {\it Kepler} space telescopes \citep{Michel08,Gilliland10,Chaplin11a}. However, few binary systems have solar-like oscillations detected in both components.

Three systems exist in which the two components have been observed separately: $\alpha$~Cen~A \citep[e.g][]{Bouchy02,Bedding04,Fletcher06,Bazot07}, and~B \citep{Carrier03,Kjeldsen05}, 16~Cyg~A~and~B \citep[KIC\,12069424 and KIC\,12069449;][]{Metcalfe12,Metcalfe15,Davies15}, and HD\,176071 \citep[KIC\,9139151 and KIC\,9139163;][]{Appourchaux12}. Additionally, oscillations in both components of HD\,177412 (HIP\,93511) have been measured while observed as a single {\it Kepler} target, KIC\,7510397 \citep{Appourchaux2015}.

More commonly, oscillations have been detected in one member of a binary system \citep[e.g.][]{Mathur13}, particularly amongst the many red giants that have been observed \citep[e.g.][]{Hekker10b,Frandsen13,Gaulme13,Gaulme14,Beck14,Rawls16}. In addition, KIC\,4471379 \citep{Murphy14} is a binary in which both components are $\delta$\,Scuti pulsators, while KIC\,10080943 \citep{Schmid15,Keen15} has two $\gamma$\,Doradus/$\delta$\,Scuti hybrids.

{\it Kepler} has detected oscillations in hundreds of Sun-like stars \citep{Chaplin11a,Chaplin14}. With an estimated $60 \%$ of stars thought to be multiple systems, it might be expected that {\it Kepler} will have observed many unresolved binary systems, some of which should exhibit oscillations in both components. However, \citet{Miglio14} used population synthesis models to predict that such systems are rare due to the requirement that both stars have oscillations with similar amplitudes in order to be detected. For this, they must have similar luminosity-to-mass ratios \citep{Kjeldsen95}. Despite this, HD\,177412 contains unequal components. The primary star of this system is $\sim$2.25 times more luminous than the secondary, and the ranges in which their oscillation modes are excited are well-separated in frequency \citep{Appourchaux2015}. In this case, the system was sufficiently bright ($V\,=\,7.9\,$mag) that the oscillation modes of the secondary star could still be detected.

In this paper we report on another asteroseismic binary showing solar-like oscillations in both components. It corresponds to the previously-known binary HD~176465, which has been observed as a single {\it Kepler} target, KIC\,10124866\footnote{Within the {\it Kepler} Asteroseismic Science Consortium (KASC), stars have attracted nicknames that have been used by those analysing them. HD\,176465\,A~and~B are referred to as ``Luke'' and ``Leia'', respectively.}. HD~176465 was identified as a visual binary from 1902 observations as part of the Astrographic Catalogue \citep{Urban98}, and has since been observed on multiple occasions \citep{Mason01}. With a Tycho visual magnitude of 8.537, the primary, HD~176465~A is only $\sim$1.2 times more luminous than HD~176465~B ($V_T\,=\,8.674\,$mag), in line with the expectations of \citet{Miglio14}. Due to the similarity of the stars, their oscillation modes are excited at similar frequencies, complicating their analysis. In this paper we present the determination of the oscillation mode parameters and detailed asteroseismic models of both HD~176465~A~and~B.

%__________________________________________________________________

\section{Observations}

\begin{figure*}
\centering
\resizebox{\hsize}{10cm}{\includegraphics{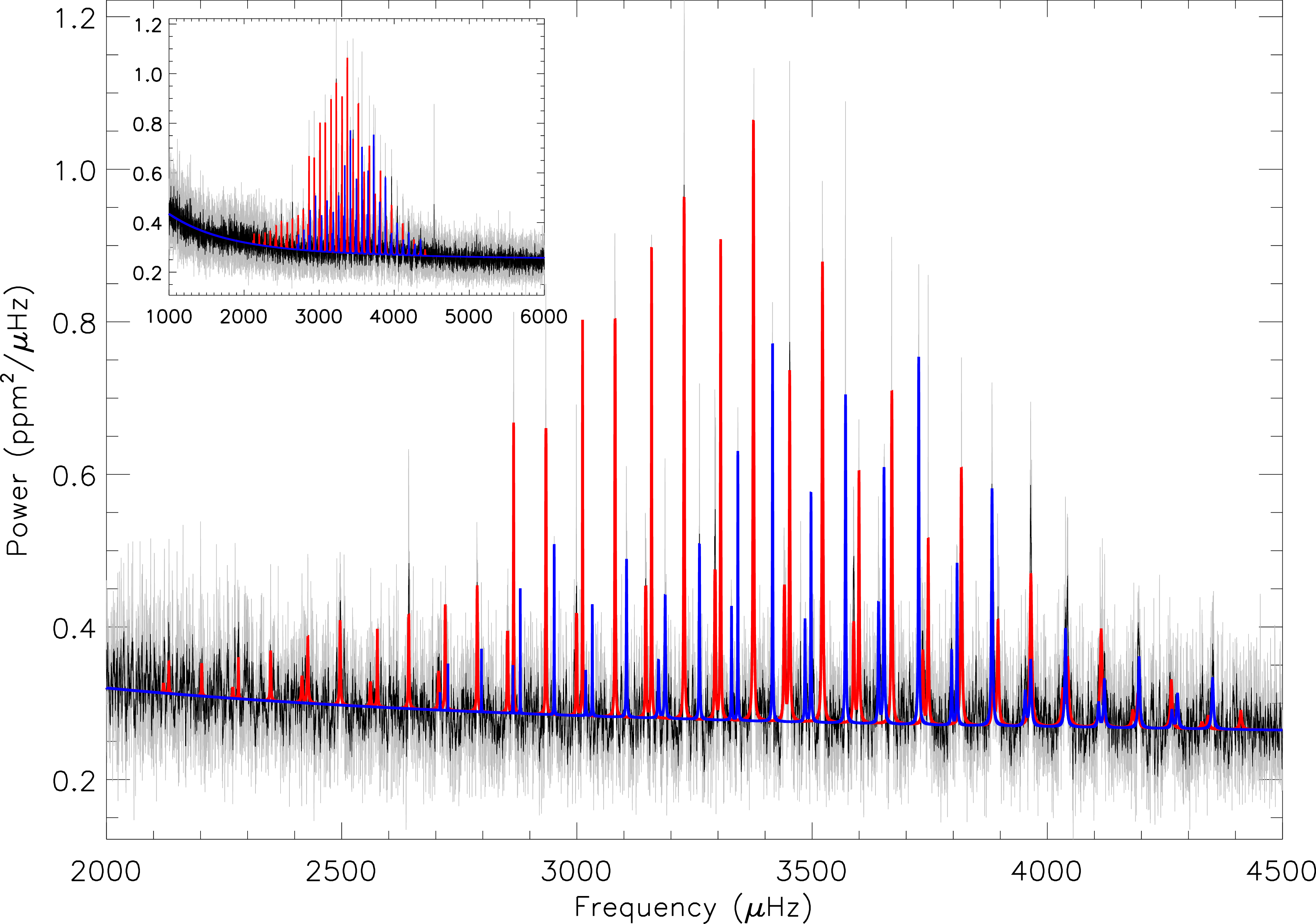}}
\caption{Power spectrum of HD\,176465, smoothed by a box-car filter over  0.5\,$\mu$Hz (grey) and 2\,$\mu$Hz (black). Superimposed is the best fit of HD\,176465A (red) and HD\,176465B (blue). The inset shows the power spectrum over a wider range. The peak at 4531\,$\mu$Hz is an artifact (see Fig.~\ref{fig:ech}).}
\label{fig:pow}
\end{figure*}

HD\,176465 was observed by the {\it Kepler} space telescope in short-cadence mode (SC; 58.85\,s sampling) for 30 days
during the asteroseismic survey phase (20 Jul 2009 to 19 Aug 2009, Q2.2), and continuously after the end of the survey (37 months, Q5--Q17).
Additionally, the system was observed in the long-cadence mode (LC; 29.43\,min sampling) for the entire nominal mission (4\,years, Q0--Q17), although this sampling is not rapid enough to sample frequencies of the solar-like oscillations, which are well above the LC Nyquist frequency. The SC time series were prepared from the raw observations as described by \citet{Jenkins10} and further corrected to remove outliers and jumps as described by \citet{Garcia11}. Figure~\ref{fig:pow} shows the smoothed
power spectrum of the short-cadence time series.

The signature of solar-like oscillations is a broad, Gaussian-like hump of power comprised of approximately equally-spaced modes.
The frequency of maximum oscillation power is denoted $\nu_\mathrm{max}$, and the oscillation frequencies are well approximated
by the asymptotic relation \citep{Vandakurov67,Tassoul80,Gough86}
\begin{equation}
\nu_{n,l} \approx \Delta\nu\left(n+\frac{l}{2}+\epsilon\right) - \delta\nu_{0l},
\label{eqn:asympt}
\end{equation}
where $\Delta\nu$, called the large separation, is the spacing between modes of the same spherical degree $l$ and consecutive radial orders $n$,
$\delta\nu_{0l}$ is the small separation between modes of different degree, and $\epsilon$ is a dimensionless phase offset. 
The presence of solar-like oscillations in both stars is clear in Fig.~\ref{fig:pow}.

\begin{figure}
\centering
\resizebox{\hsize}{!}{\includegraphics{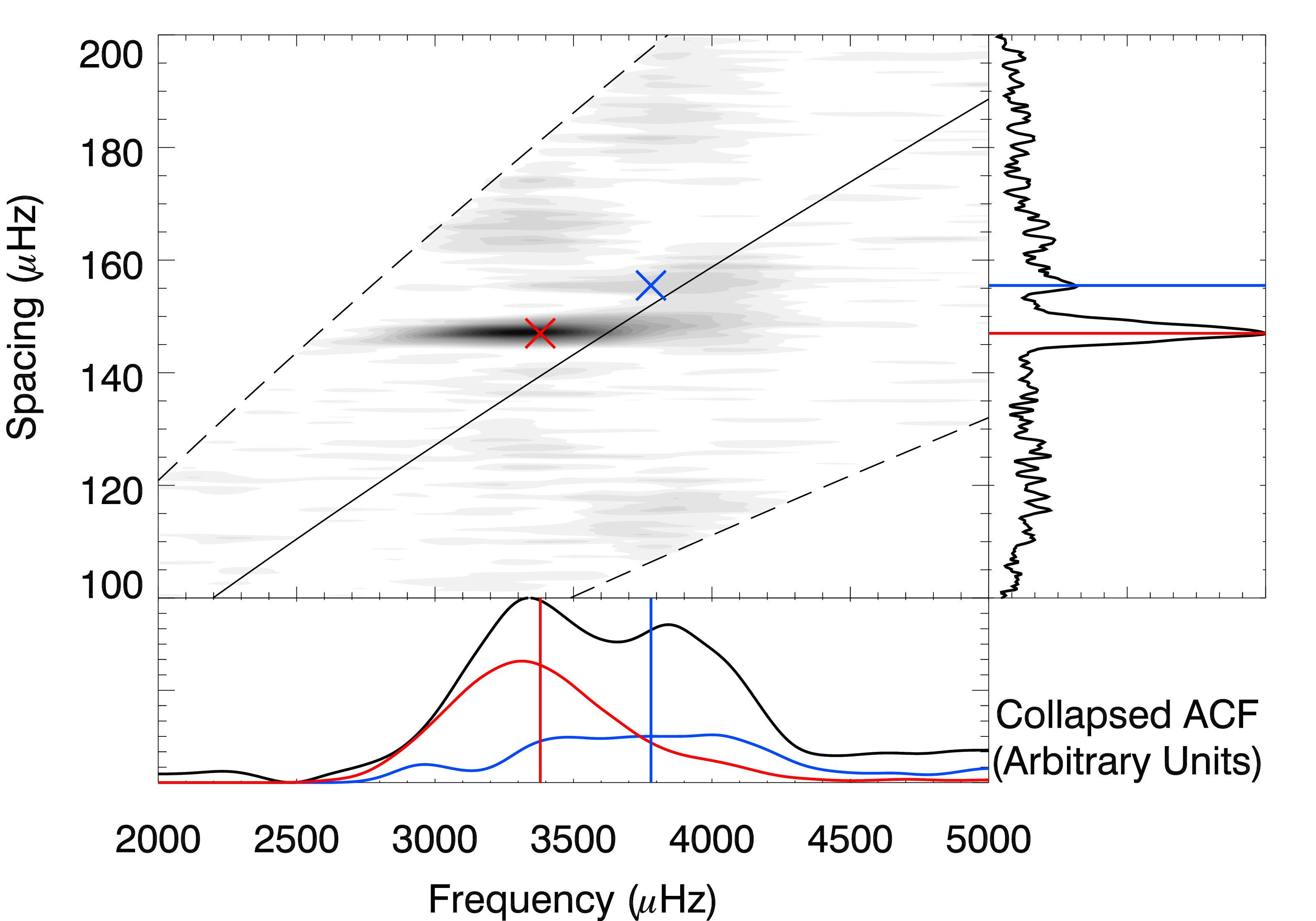}}
\caption{Localized autocorrelation of the power spectrum of HD\,176465. Two peaks in the autocorrelation function are visible, corresponding to the primary and secondary components of HD\,176465 (red and blue crosses, respectively). The solid line shows the mean observed relation between $\Delta\nu$ and $\nu_\mathrm{max}$ \citep{Stello09}, while the dashed lines are 0.7~and~1.3 times this value. The collapsed autocorrelation function (ACF) along each axis are shown in black in the lower and right panels, with the peaks indicated by red and blue lines. The red and blue curves in the lower panel show the collapsed ACF with spacings around 147 and 155 $\mu$Hz, respectively.}
\label{fig:acf}
\end{figure}

Oscillations were detected in the survey data (Q2.2), and the global oscillation parameters $\Delta\nu$ and 
$\nu_{\mathrm{max}}$ were extracted using several automated pipelines \citep{Campante10,C-D08c,Hekker10,Huber09,Karoff10,
Mathur10,Mosser09,Roxburgh09}. Inspection of the autocorrelation function of the power spectrum, shown in Fig.~\ref{fig:acf},
reveals a strong peak that corresponds to the higher-amplitude oscillations of HD~176475~A. While most pipelines were able to make this detection, as they were programmed to seek oscillations from a single star, the presence of a second set of oscillations in the time series was missed.

\begin{figure*}
\centering
\resizebox{\hsize}{!}{\includegraphics{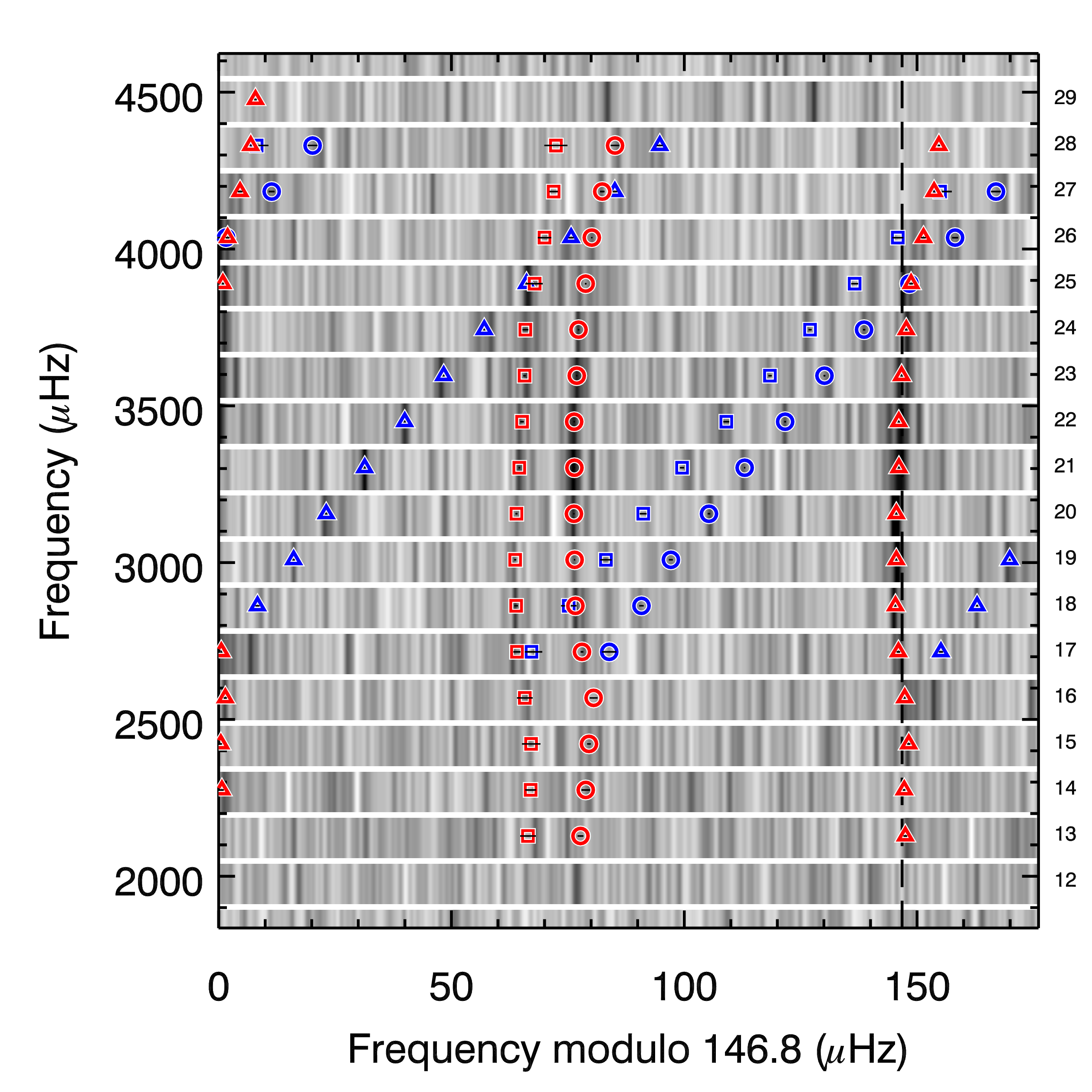}\includegraphics{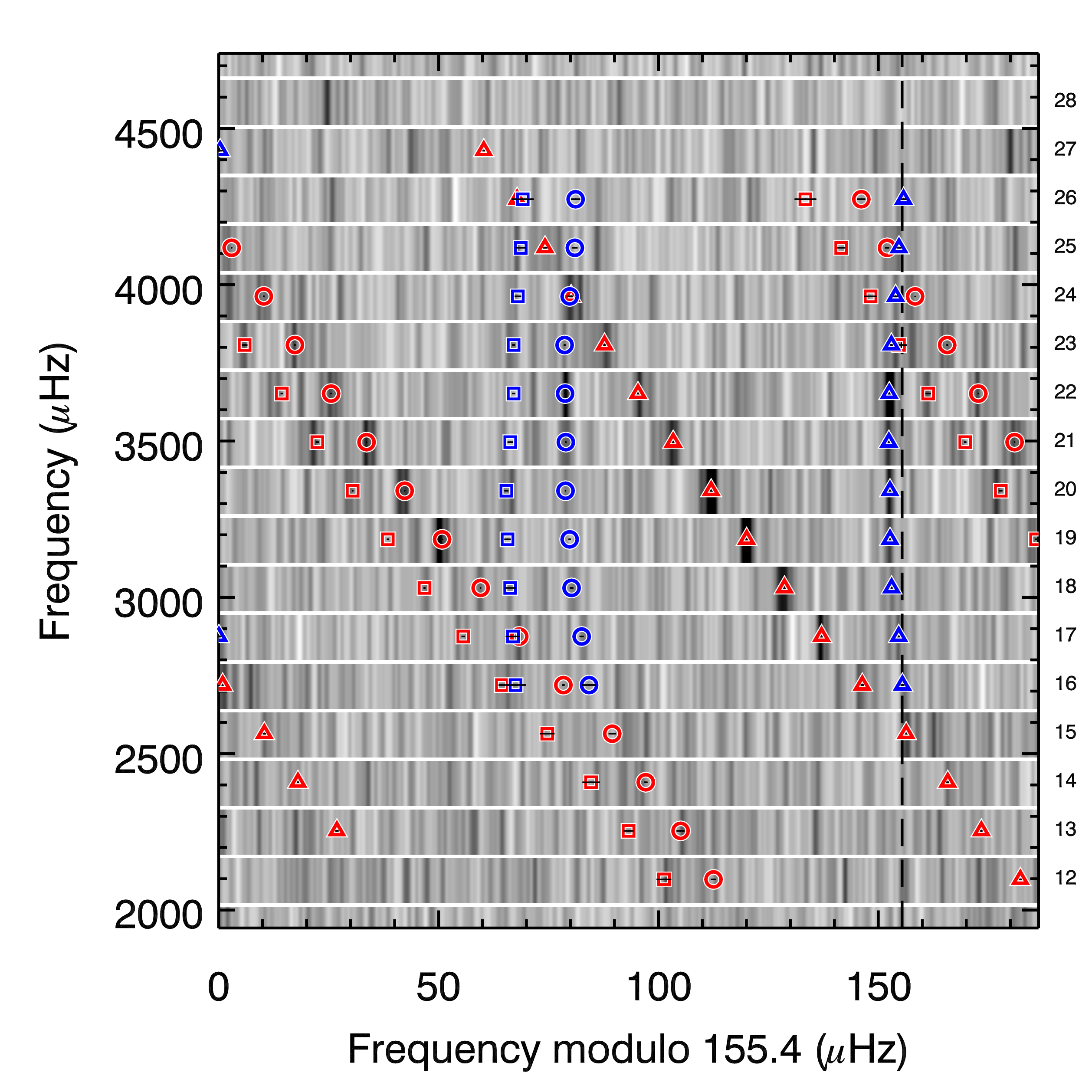}}
\caption{\'Echelle diagrams of HD\,176465\,A (left)~and~B (right) showing the overlapping sets of modes with different large separations. The fitted frequencies are indicated by the red and blue symbols for the A~and~B components respectively. Symbol shapes indicate the mode identification: circles are $l=0$ modes, triangles are $l=1$ modes, and squares are $l=2$ modes. For reference, a greyscale map of the smoothed power spectrum is shown in the background. Numbers to the right of the plot indicate radial order of the $l=0$ modes. To make the location of modes near the edge of the \'echelle diagrams clearer, the diagrams have been extended in width beyond the value of the large separation, which is indicated by the dashed vertical lines. The peak at 4531\,$\mu$Hz is the 8th harmonic of the {\it Kepler} long-cadence sampling rate, which is a known artifact \citep{Gilliland10b}.}
\label{fig:ech}
\end{figure*}

The realization that the oscillations of this second star were present in the {\it Kepler} timeseries was made by a manual inspection
of the \'echelle diagram. Figure~\ref{fig:ech}~(left) shows the power spectrum in \'echelle format with a
large separation corresponding to that of the primary, $\Delta\nu=146.8\,\mu$Hz. 

Near-vertical ridges in the \'echelle diagram (red symbols) correspond to modes of various radial orders, $n$, and the same spherical degree, $l$. The closely spaced $l=2,0$ ridges and the $l=1$ ridges are clearly visible, from left to right. We also see other modes that do not fall along vertical ridges in this \'echelle diagram, but instead form sloping ridges (blue symbols).

A diagonal ridge in an \'echelle diagram means that the large separation is incorrect for those modes. In this case, the value of $\Delta\nu$ is too small but for a larger value, these ridges will become near-vertical. Figure~\ref{fig:ech}~(right) shows the \'echelle diagram with
$\Delta\nu=155.4\,\mu$Hz. The ridges that were diagonal in Fig.~\ref{fig:ech}~(left) have now become vertical, and the 
formerly vertical ridges are now diagonal, but sloping downwards. 

Since the large separation approximately scales with the square root of the mean stellar density \citep{Ulrich86}, the existence of two distinct values of $\Delta\nu$ indicate the presence of two distinct stars. We can be confident that the oscillations with $\Delta\nu=146.8\,\mu$Hz are those of HD~176465~A, and oscillations with $\Delta\nu=155.4\,\mu$Hz are those of HD~176465~B because, for a given age, more massive stars have a lower mean density \citep{C-D84}.

%______________________________________________________________

\section{Methods to measure the mode parameters}
Extracting accurate mode parameters is an important step required before modelling a star. The most common method is the Maximum Likelihood Estimator (MLE) approach \citep{Anderson1990}, which has been extensively used to analyse the low-degree global acoustic oscillations of the Sun \cite[e.g.][]{Chaplin1996}. 
For solar-like oscillations, the noise statistics for the power spectral density $y_i$ at a given frequency $\nu_i$ is a $\chi^2$ with two degrees of freedom so that the likelihood is
\begin{equation}
	L=\prod_{i=1}^N\frac{1}{M(\nu_i, \boldsymbol{\theta})} \mathrm{e}^{- y_i/M(\nu_i, \boldsymbol{\theta})}.
\end{equation}
Here, $M(\nu_i, \theta)$ is the model used to describe the data set $\{\nu_i, y_i\}$ and $\boldsymbol{\theta}$ are the parameters of that model. The model of the power spectrum has two components. Firstly, the noise background is comprised of a sum of quasi-Lorentzian functions whose parametrization may vary \citep{Harvey1985, Karoff2012} plus a flat white noise level. Secondly, the acoustic p modes are described as a series of Lorentzian profiles, with frequencies, heights, and linewidths to be determined.

The MLE approach is well-suited to cases where the likelihood function has a well-defined single maximum, so that convergence towards an unbiased measure of the fitted parameters is ensured \citep{Appourchaux98}. In the case of the Sun, this is often the case, mostly because of the very high signal-to-noise of the pulsation modes. However, stellar pulsations often have much lower signal-to-noise ratios, so that the likelihood may have several local maxima. In such a case, MLE methods may not converge towards the true absolute maximum of probability.

A Bayesian approach was proposed to tackle this problem \citep{Benomar09b, Benomar09, Handberg11,Corsaro14}. The Bayesian approach uses conditional probabilities and involves Bayes' theorem to define a so-called posterior distribution $\pi(\boldsymbol{\theta}, \boldsymbol{y} | M)$. This statistical criterion incorporates not only the likelihood $L \equiv  \pi(\boldsymbol{y}|\boldsymbol{\theta}, M)$, but also the {\it a priori} information $\pi(\boldsymbol{\theta} | M)$ on the fitted parameters,
\begin{equation}
	\pi(\boldsymbol{\theta}, \boldsymbol{y} | M) = \frac{\pi(\boldsymbol{\theta} | M) \pi(\boldsymbol{y}|\boldsymbol{\theta}, M)}{\pi(\boldsymbol{y}|M)}.
\end{equation}
Here, $\pi(\boldsymbol{y}|M)$ corresponds to the global likelihood, linked to the posterior probability of the model given the data,
\begin{equation} \label{eq:posterior_model}
	P(M | \boldsymbol{y})=\frac{\pi(\boldsymbol{y}|M) \pi(M)}{\pi(\boldsymbol{y})},
\end{equation}
which is the probability that a model $M$ is consistent with observations $\boldsymbol{y}$. This is an essential quantity for comparing competing models, since it defines the level of {\it evidence} supporting each one of the models. 
In its most sophisticated and useful form, the Bayesian method uses a sampling algorithm, such as the Markov Chain Monte Carlo (MCMC) sampler. This is actually crucial to calculating the global likelihood\footnote{The global likelihood requires integration over the parameters $\boldsymbol{\theta}$, which is often impossible to carry out analytically.}. 
By exploring the full extent of the posterior distribution (within the bounds defined by the priors), one can determine the full probability distribution of each of the parameters. Although it is significantly more computationally intensive than the MLE, this approach is less sensitive to the potential local maxima, because they are all sampled\footnote{This is a limit property, i.e. when the number of samples tend towards infinity. Due to the finite number of samples, in practice it is not certain that all of the parameter space is explored.}.
Note that the Bayesian method does not necessarily involve a sampling method. In its less sophisticated form, a simple maximisation (in contrast to a full sampling) of the posterior probability is performed. This is the  Maximum A Posteriori (MAP) approach. This regularized maximization, in principle, outperforms the MLE in the case of informative priors, but could still be sensitive to local maxima of probability and does not enable the calculation of the evidence. 

\section{Mode Parameters} \label{sec:mode_param}

In contrast to the binary star system HD~177412 \citep{Appourchaux2015}, the extraction of mode parameters from the power spectrum of HD\,176465 is complicated by the overlapping frequencies of the two stars. While we have followed the fitting strategy used in previous analyses of single stars 
\citep[e.g.][]{Campante11,Mathur11,Mathur13,Appourchaux12}, extra care was required to ensure a correct fit of the power spectrum. 

Initial values for the mode frequencies were obtained by scaling the solar p-mode frequencies measured by BiSON \citep{Broomhall09}. 
The solar frequencies were scaled by the ratio of the large separations of HD\,176465 to the Sun \citep{Bedding10b}. To account for the different values of the small separations $\delta\nu_{0l}$ and phase offset $\epsilon$ in these stars compared to the Sun, uniform shifts were then applied to all modes of the same degree so that
the scaled and shifted solar frequencies corresponded to the observed peaks in the power spectrum of HD\,176465.

Several teams then used these estimates as inputs to perform a comprehensive fit of the power spectrum, including the background, for both stars
simultaneously. The power spectrum of each star was modelled as a sum of Lorentzian profiles, with frequencies $\nu_{n,l}$, heights $H$ and linewidths $\Gamma$ as free parameters.
Additionally, some teams also included the effect of rotation, with the rotational splitting frequency $\nu_s$ and inclination of the stellar rotation axis $i$ as additional free parameters.

The teams differed in how they approached finding the best fit, with some using MLE \citep[e.g.][]{Appourchaux98}, others using the MAP method 
\citep[e.g.][]{Gaulme09}, or the MCMC method \citep[e.g.][]{Benomar08}.

To identify a single set of mode parameters to be adopted in our analysis, the frequencies reported by each team were compared to find the ``best'' fitter that provided
the smallest deviation from the average frequencies across the power spectrum. Statistical outlier rejection was performed using Peirce's criterion \citep{Peirce52,Gould55}, as previously used in the analysis of other Kepler targets \citep{Campante11,Mathur11,Appourchaux12}. Care was required where the frequencies of the two stars overlapped to ensure that the modes were identified with the correct star.

The ``best'' fitter, which used a Bayesian approach coupled with a MCMC algorithm, then performed a final analysis of the power spectrum. The two stars were fitted simultaneously in order to separate their overlapping stellar pulsations (see Fig.~\ref{fig:pow}). A local fit was then performed to determine the probability of detection for each pulsation mode. A thorough description of the method and of the priors is given in Appendix \ref{app:A}.

\begin{figure*}
\centering
\resizebox{\hsize}{!}{\includegraphics{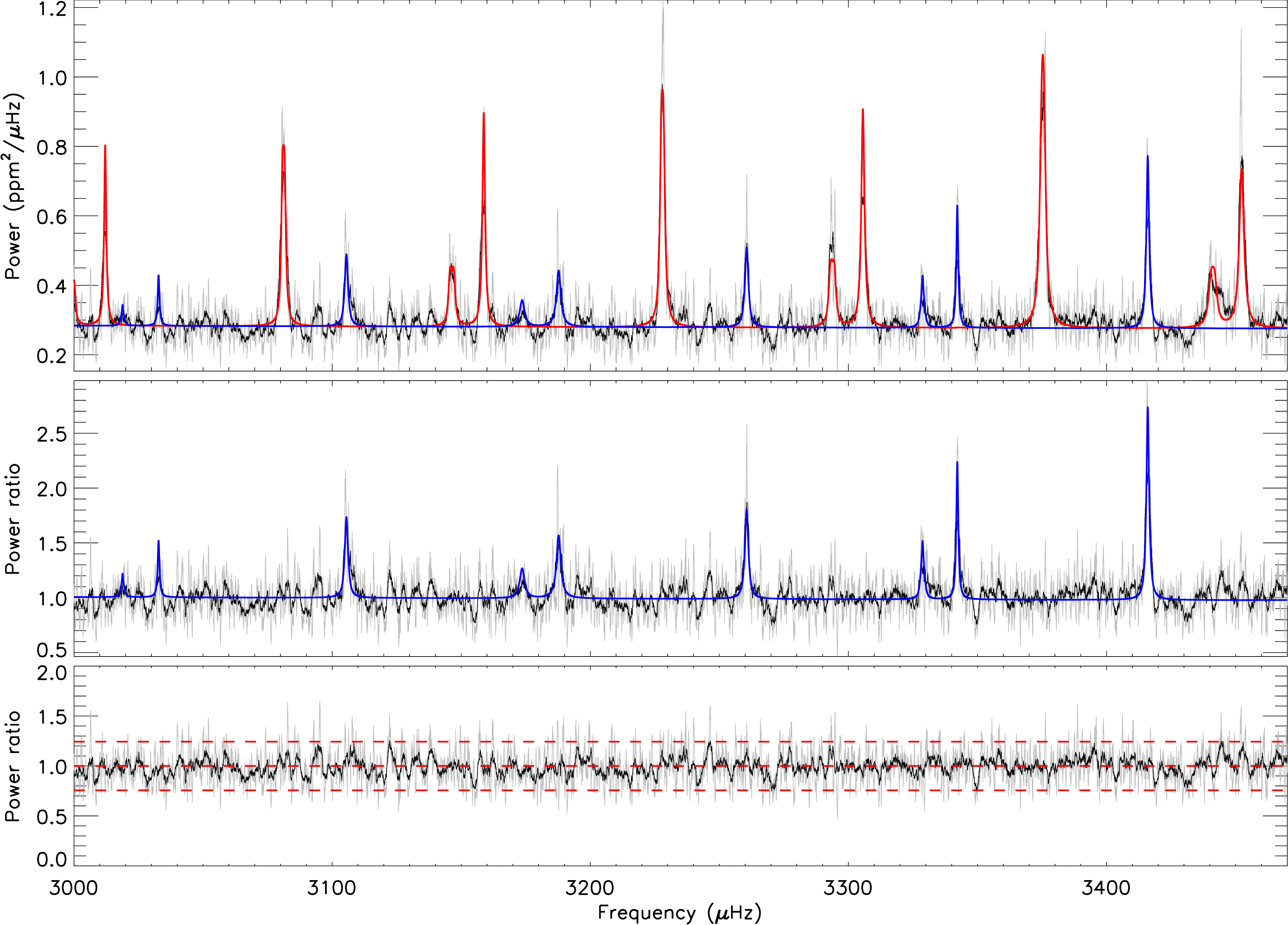}}
\caption{{Top.} Smoothed power spectrum using a boxcar filter over 0.5 $\mu$Hz (grey) and 2$\mu$Hz (black) between 3000 $\mu$Hz and 3470 $\mu$Hz. Superimposed is the best fit for HD 176465 A (red) and HD 176465 B (blue). Middle. Residual power spectrum of HD 176465 B, obtained by dividing the original spectrum with the best fit of HD 176465 A. Bottom. Residual power spectrum obtained by dividing the original spectrum with the best fit for HD 176465 A and B. Data points between the lower/higher red-dashed lines have $95\%$ probability to be due to noise for the black power spectrum.}
\label{fig:fit:goodness}
\end{figure*}

The fitted mode parameters are given in Table~\ref{tab:Luke_freq} for HD\,176465\,A and Table~\ref{tab:Leia_freq} for HD\,176465\,B. The fitted frequencies are shown in the \'echelle diagrams in Fig.~\ref{fig:ech}. We show in Fig.~\ref{fig:fit:goodness} the residual power spectrum after removing the signal from HD\,176465\,A (middle panel) and from both stars (bottom panel). The lack of significant residual power indicates the precision of this fit.

\begin{figure*}
\centering
\resizebox{\hsize}{!}{\includegraphics[angle=0, trim = -5mm -5mm -5mm -5mm, clip]{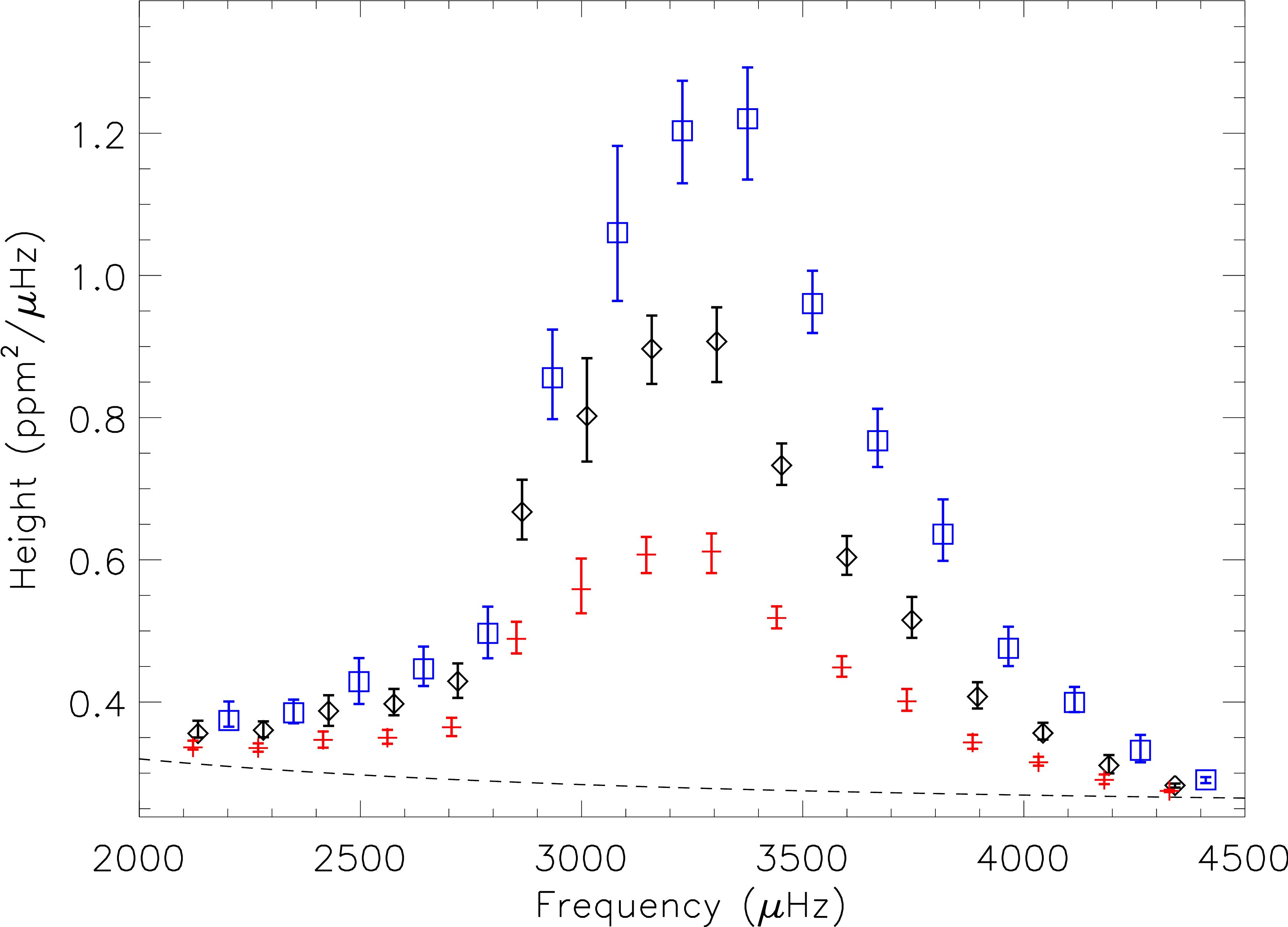}\includegraphics[angle=0, trim = -5mm -5mm -5mm -5mm, clip]{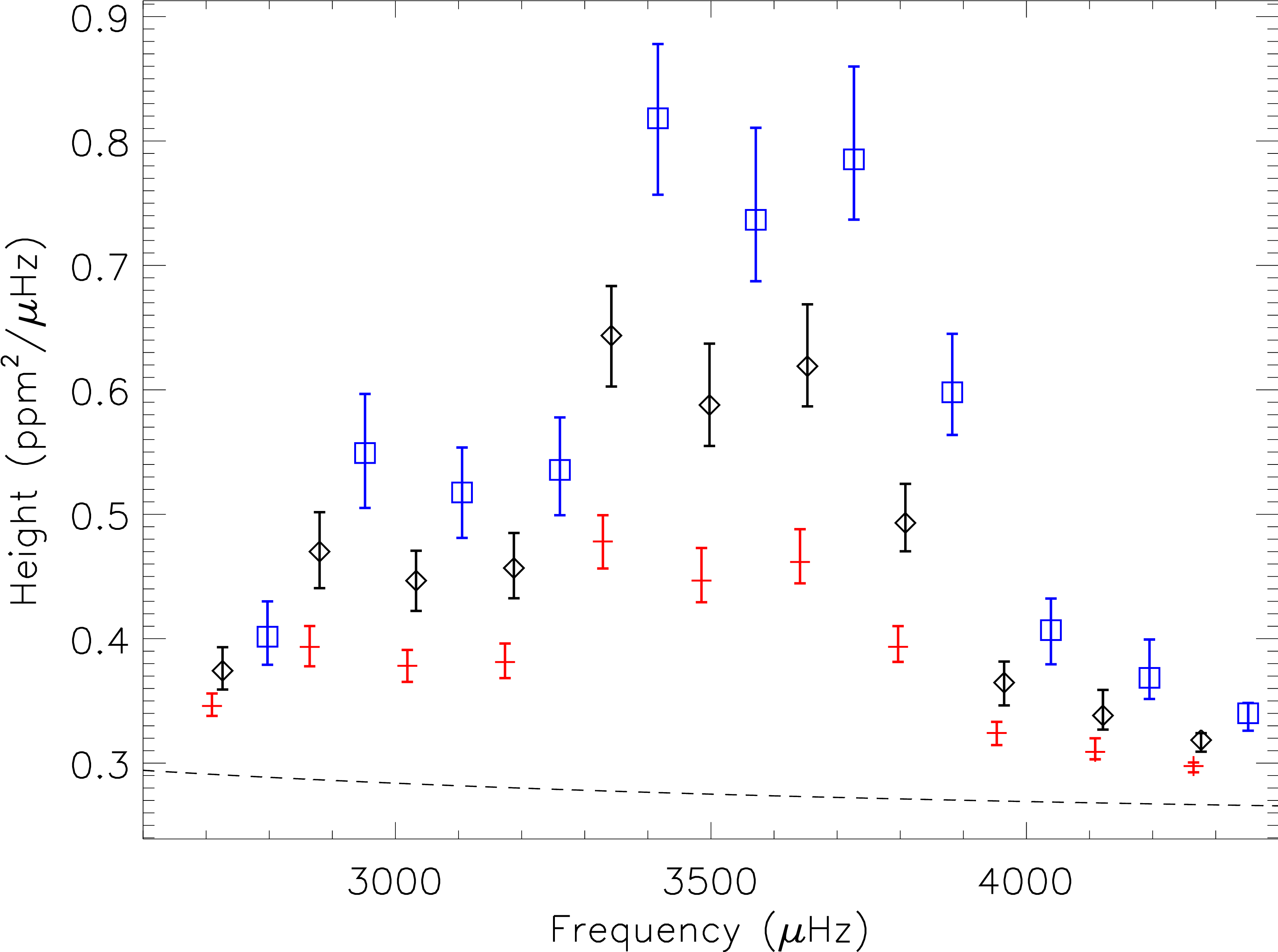}}
\caption{Mode heights for HD\,176465\,A (left) and B (right). Angular degree, $l=0,1,2$, of the modes are indicated by the black diamonds, blue squares, and red crosses, respectively. The dashed lines indicate the noise background.}
\label{fig:heights}
\end{figure*}

\begin{figure*}
\centering
\resizebox{\hsize}{!}{\includegraphics[angle=0, trim = -5mm -5mm -5mm -5mm, clip]{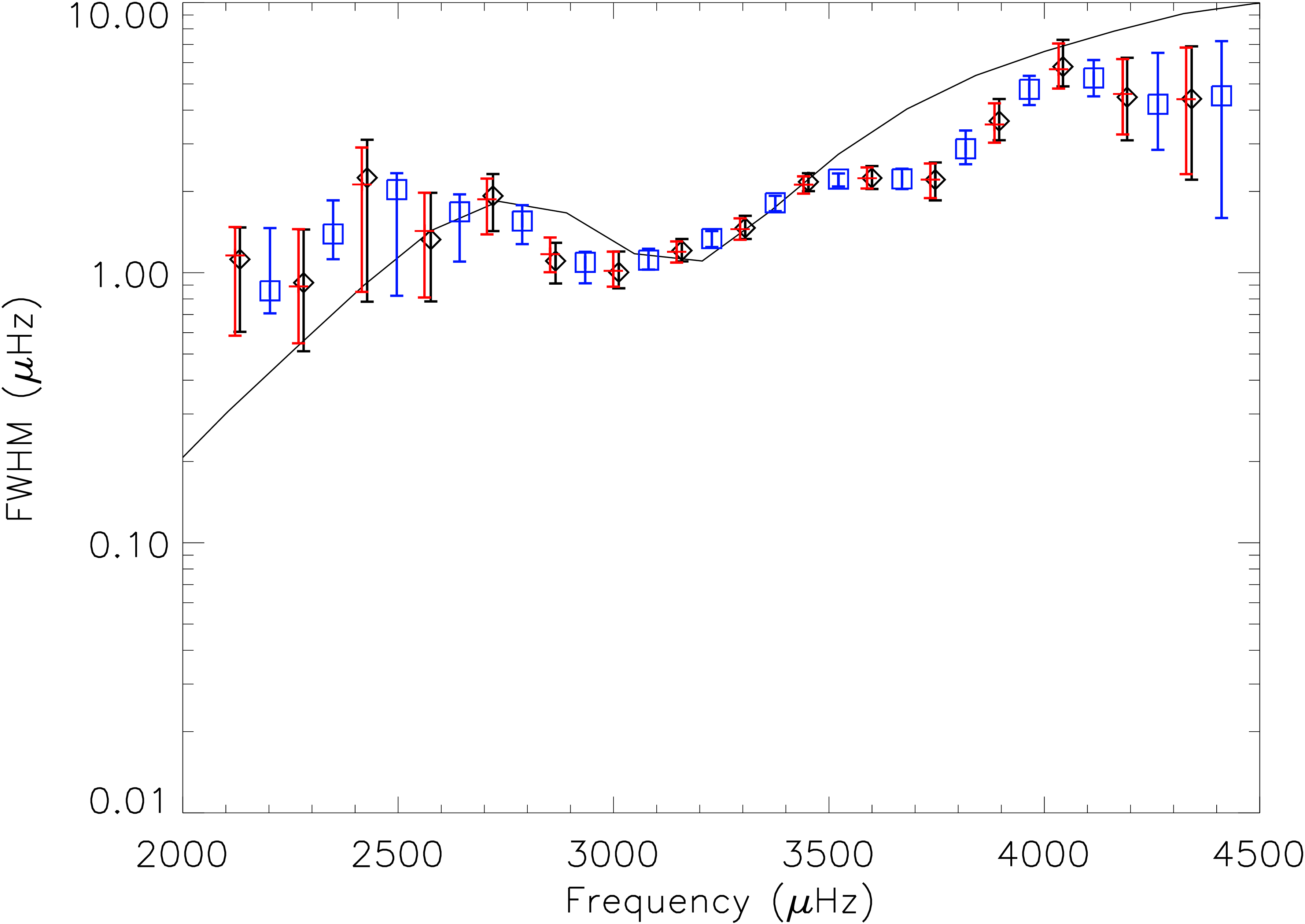}\includegraphics[angle=0, trim = -5mm -5mm -5mm -5mm, clip]{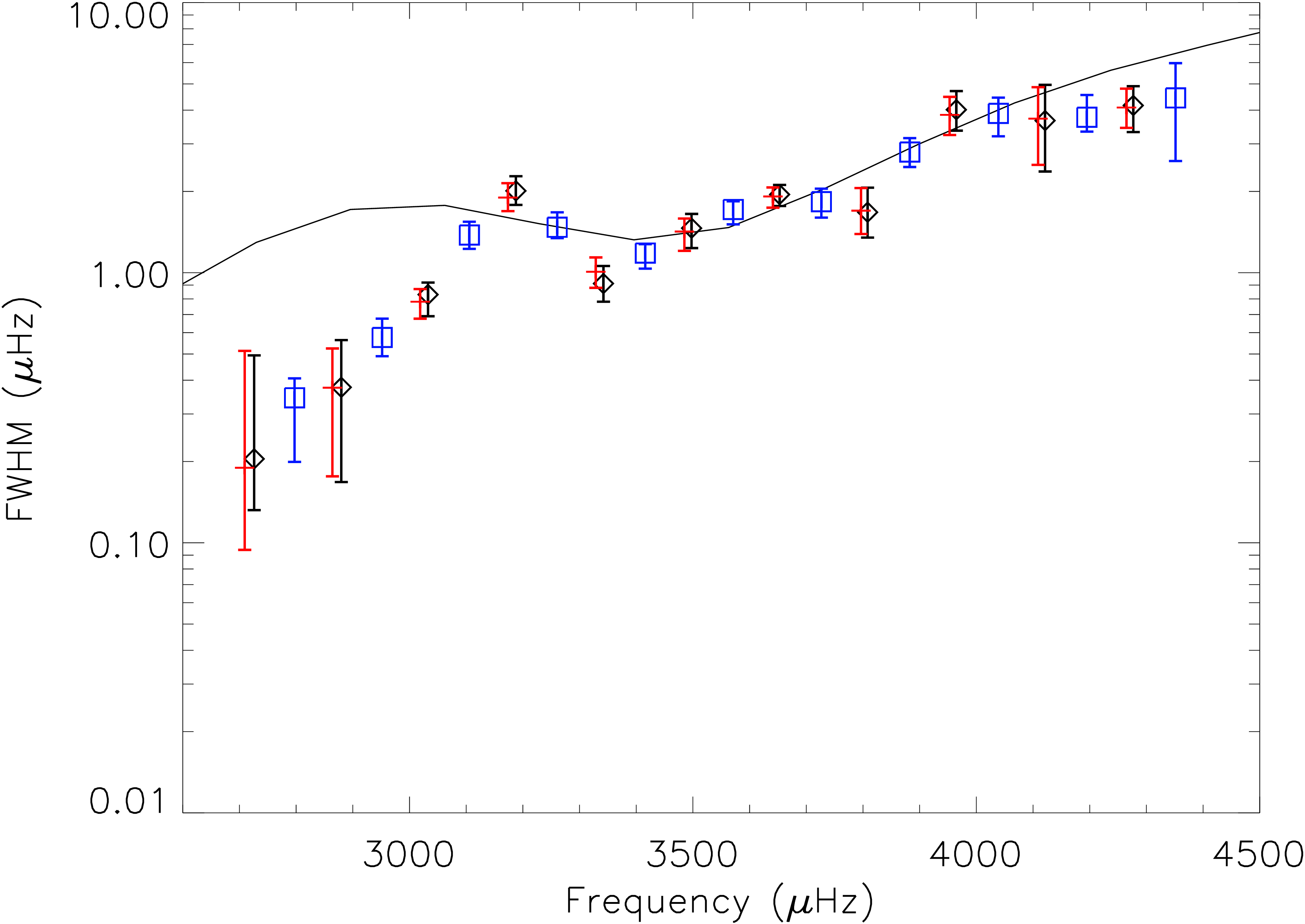}}
\caption{Mode widths for HD\,176465\,A (left) and B (right). Angular degree, $l=0,1,2$, of the modes are indicated by the black diamonds, blue squares, and red crosses, respectively. The solid curves show the theoretical linewidths (Houdek et al. 1999; Houdek et al., in preparation) computed from the ASTFIT models discussed in Sect.~\ref{sec:model}.}
\label{fig:widths}
\end{figure*}

\begin{figure*}
\centering
\resizebox{\hsize}{!}{\includegraphics[angle=0, trim = -5mm -5mm -5mm -5mm, clip]{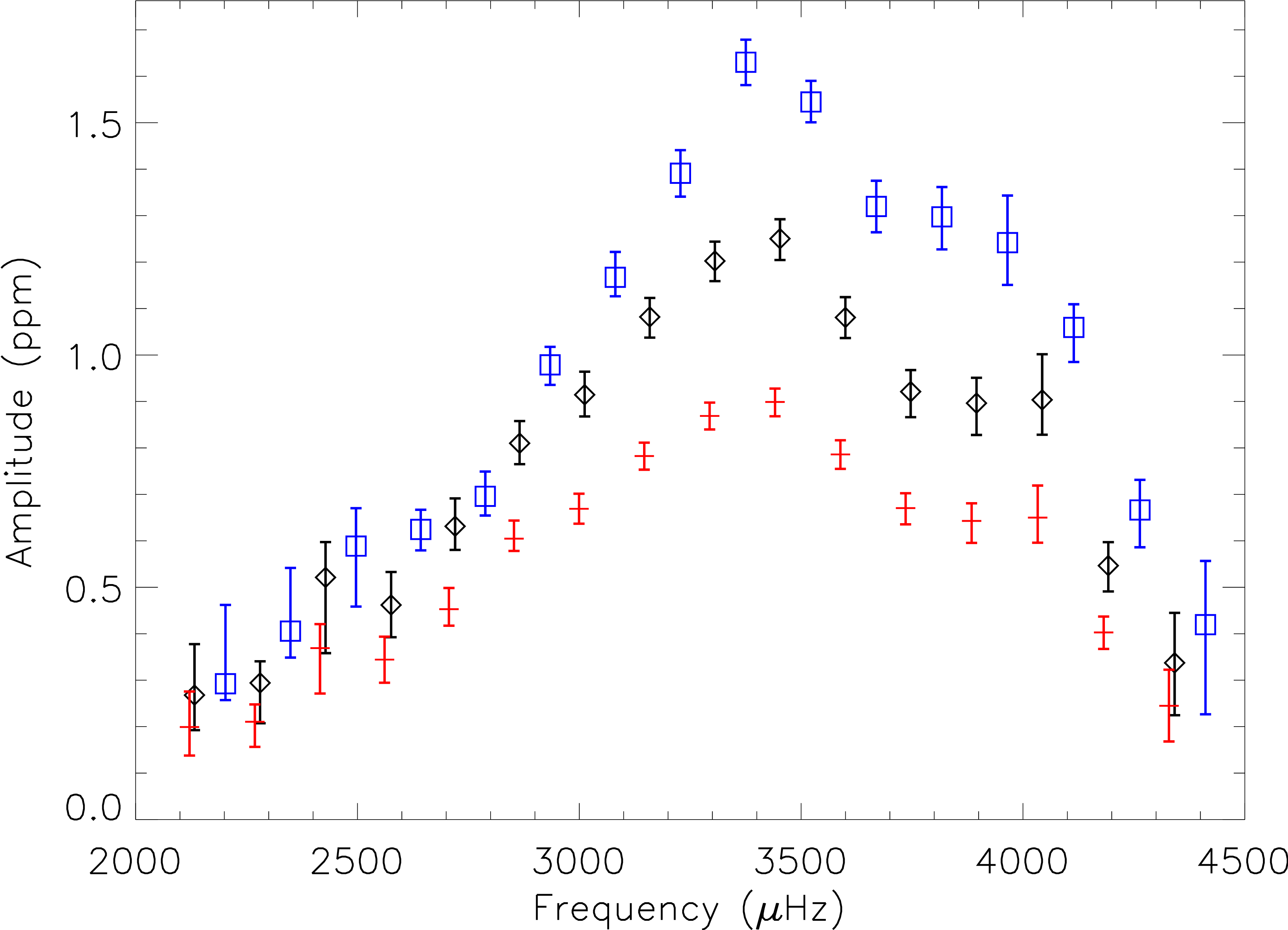}\includegraphics[angle=0, trim = -5mm -5mm -5mm -5mm, clip]{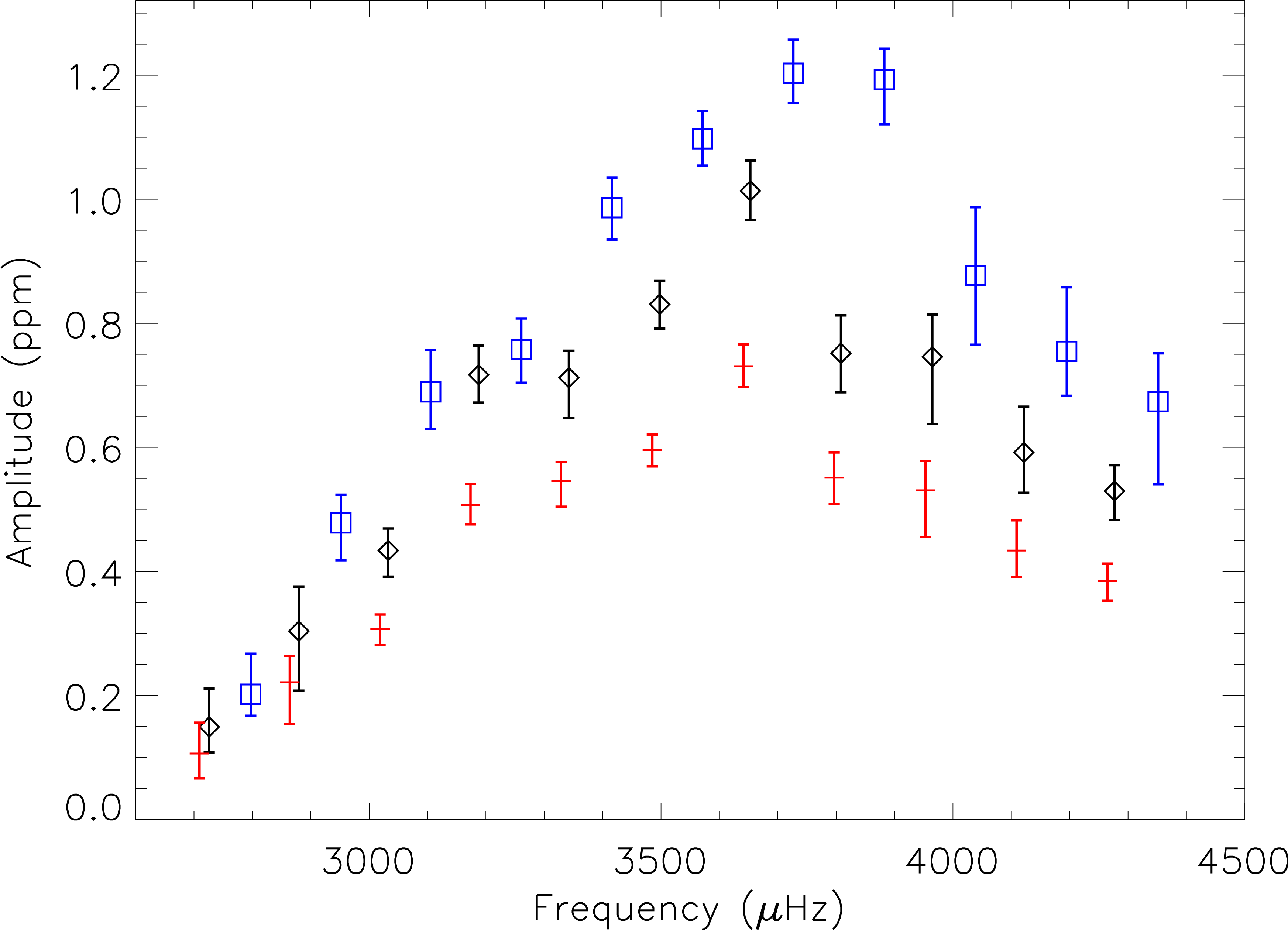}}
\caption{Mode amplitudes for HD\,176465\,A (left) and B (right). Angular degree, $l=0,1,2$, of the modes are indicated by the black diamonds, blue squares, and red crosses, respectively. These are defined as $\sqrt{\pi H \Gamma}$.}
\label{fig:amplitudes}
\end{figure*}

Mode heights are shown in Fig.~\ref{fig:heights} and linewidths in Fig.~\ref{fig:widths}. Figure~\ref{fig:amplitudes} shows the mode amplitudes, defined as $\sqrt{\pi H \Gamma}$. The mode heights and amplitudes are lower than would be the case if these stars were observed separately  because the oscillation signals are diluted by the presence of the other star. For the A component, the heights and amplitudes are lower by a factor of 1.8; for the B component, they are lower by a factor of 2.2.

In Fig.~\ref{fig:widths}, we note that both stars have a width profile similar to the Sun, with a plateau of nearly constant linewidths around $\nu_\mathrm{max}$. As a comparison, we have computed theoretical linewidths for the best-fitting models determined using the ASTFIT method we present in Sect.~\ref{sec:model}. The theoretical linewidths are calculated as described by \citet{Balmforth92} and \citet{Houdek99}, and are shown as the solid curve in Fig.~\ref{fig:widths}. In general, there is good qualitative agreement between the observed and theoretical linewidths. In HD\,176465\,B, however, the observed linewidth drops off sharply below $\sim$3200 $\mu$Hz. As a likely explanation, we note that \cite{Appourchaux14} found widths and heights may not be accurate for low signal-to-noise modes and the Lorentzian tails of the nearby modes from the primary component may have biased the noise estimate.

\begin{figure*}
\centering
\resizebox{\hsize}{!}{\includegraphics[angle=0, trim = -2mm -2mm -2mm -2mm, clip]{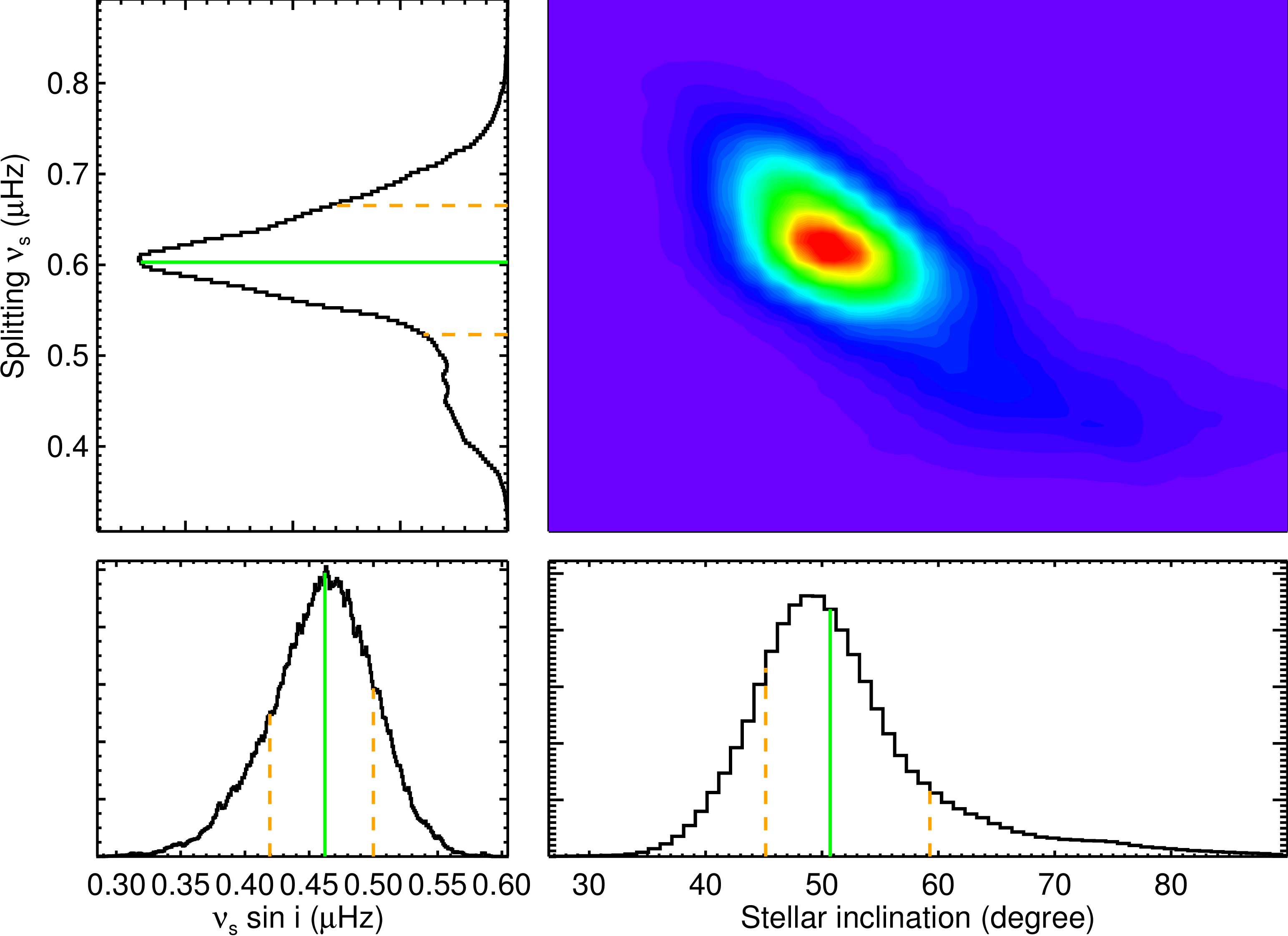}\includegraphics[angle=0, trim = -2mm -2mm -2mm -2mm, clip]{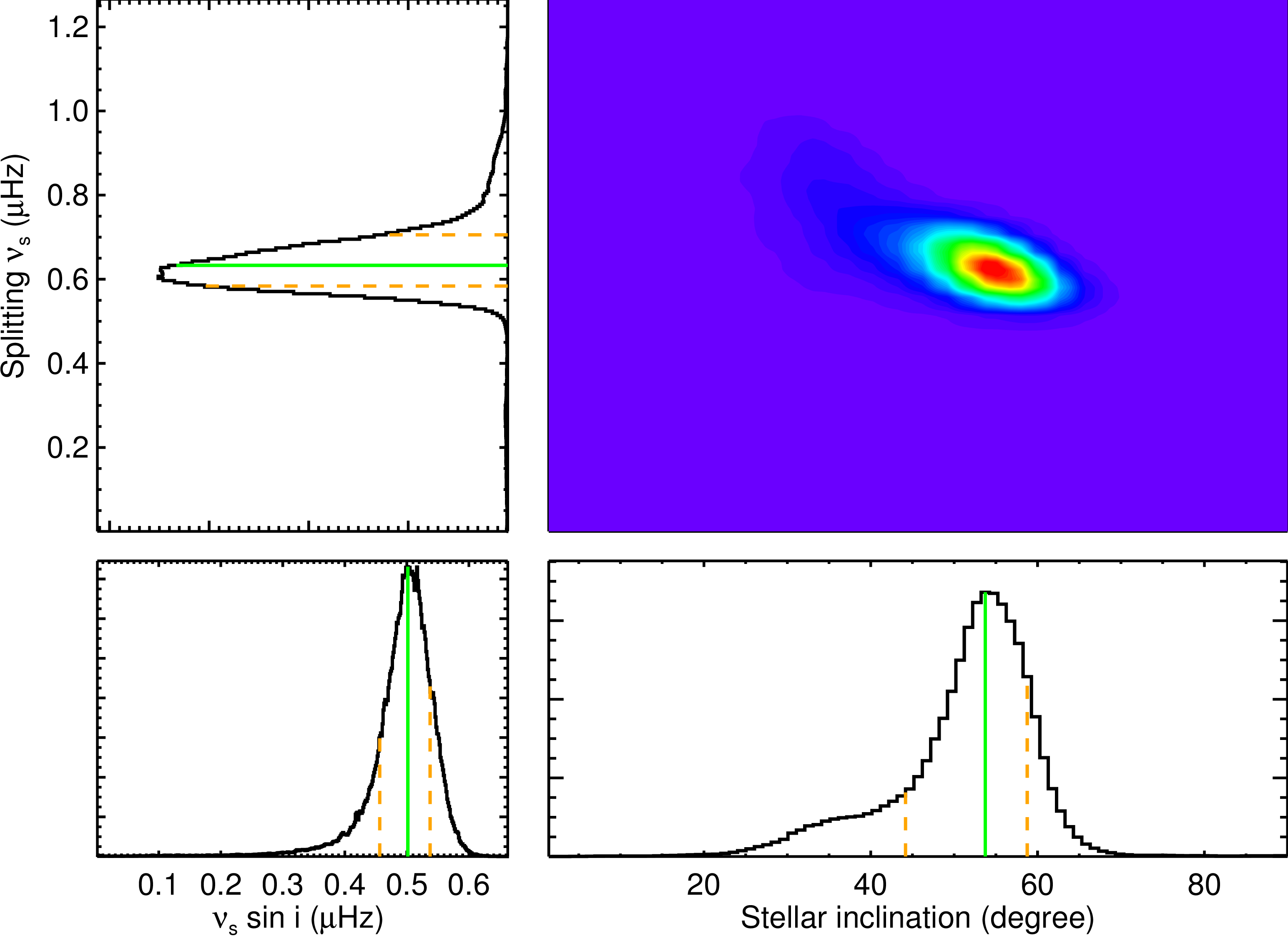}}
\caption{Posterior distribution of the frequency splitting as a function of inclination angle for HD\,176465\,A (left) and B (right). The marginalized distributions for the splitting and inclination angle are shown in the histograms along each axis. The distributions for $v_\mathrm{s} \sin i$ are shown in the lower-left corners. The solid green lines indicate the median value of the distributions, while the dashed orange lines indicate the 68\% confidence intervals.}
\label{fig:rotAB}
\end{figure*}

The results for the inclination angles and rotational splittings are given in Fig.~\ref{fig:rotAB}. The A component of the system has an internal average rotation period of 17.5--22.3\,d and has a stellar inclination of 45.2$^\circ$--59.7$^\circ$ (using confidence intervals at $1\sigma$). The B component rotation period is 16.3--19.6\,d with an inclination of 44.1$^\circ$--57.9$^\circ$. Within uncertainties, the two components therefore have the same inclination.

%__________________________________________________________________

\section{Non-Seismic Stellar Parameters}\label{sec:non_seism}

\subsection{Orbit}
\begin{figure}
\center
\resizebox{\hsize}{!}{\includegraphics{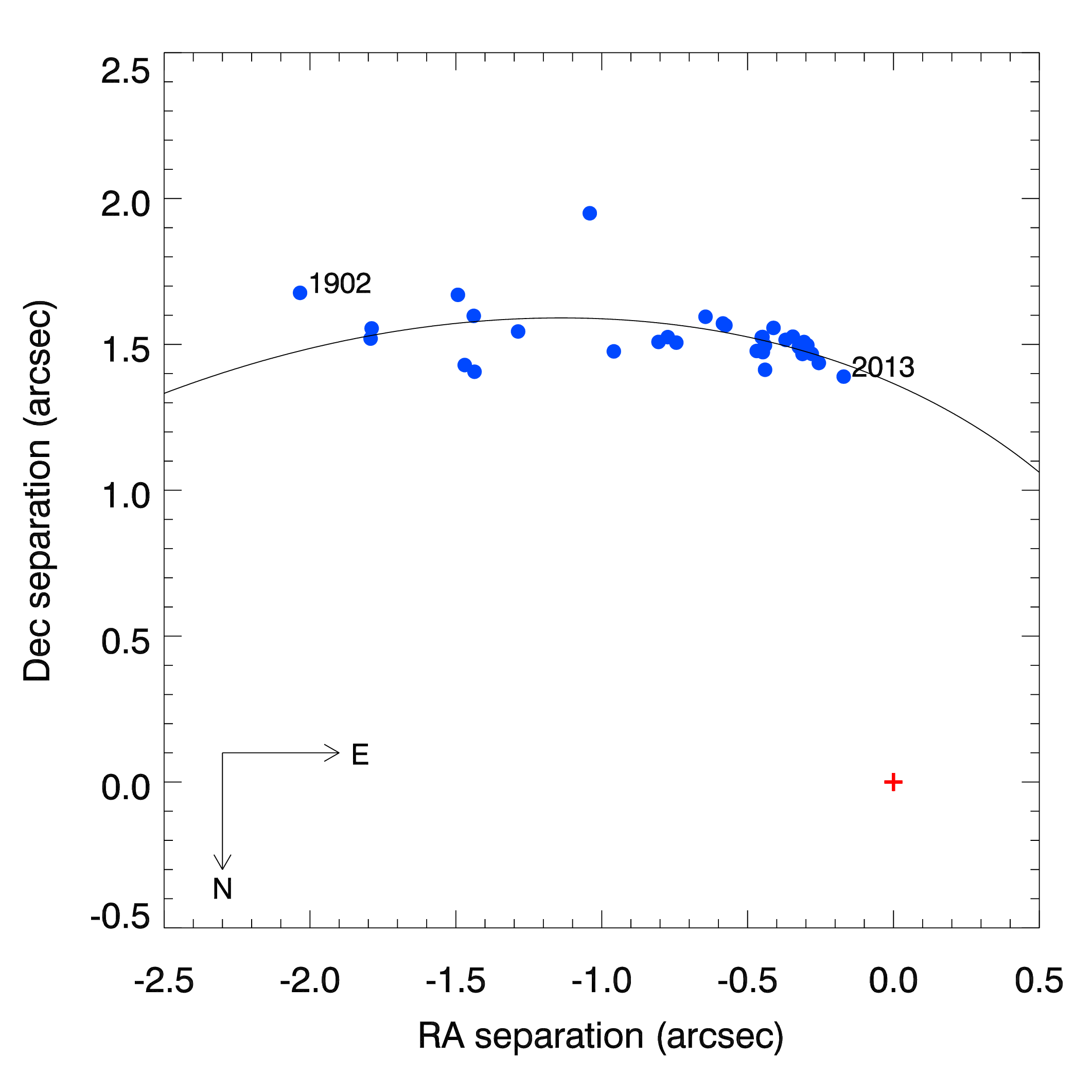}}
\caption{Positions of the secondary component (blue circles) relative to the primary (red cross) from the Washington Double Star Catalog. The curve is an ellipse fitted to the observations.}\label{fig:orbit}
\end{figure}

Position measurements in the Washington Double Star Catalog date back to 1902 \citep{Mason01}. Over the last century the position angle of the binary has 
changed by 40$^\circ$, indicating a period on the order of 1000 years. The position of the secondary relative to the primary is shown in Fig.~\ref{fig:orbit}. 
Unfortunately, an insufficient fraction of the orbit has been covered to constrain the total mass. Currently the stars are separated by $\sim$1.4". 

\subsection{Spectroscopy}
A spectrum of HD\,176465 was obtained on
01 Apr 2010 with the ESPaDOnS spectrograph on the 3.6\,m Canada-France-Hawaii Telescope. The spectrograph is fibre-fed,
and light from both stars is expected to have fallen within the fibre footprint. The stars are very similar in temperature
and metallicity, and so, unless the system is double-lined, distinguishing both stars in the same spectra would be
difficult. The spectrum has been analysed as a single star by \citet{Bruntt12} and \citet{MolendaZakowicz13} with different
spectroscopic methods. \citet{Bruntt12} used \textsc{vwa} to derive the fundamental stellar paremeters and elemental abundances from
the spectra, finding $T_\mathrm{eff}$=\,5755$\pm$60\,K, $\log g$=\,4.48$\pm$0.03\,dex and [Fe/H]=\,$-$0.30$\pm$0.06\,dex. 
\citet{MolendaZakowicz13} used \textsc{rotfit}, finding $T_\mathrm{eff}$=\,5736$\pm$92\,K, $\log g$=\,4.29$\pm$0.21\,dex and [Fe/H]=\,$-$0.31$\pm$0.21\,dex,
and \textsc{ares+moog}, finding $T_\mathrm{eff}$=\,5864$\pm$68\,K, $\log g$=\,4.57$\pm$0.11\,dex and [Fe/H]=\,$-$0.24$\pm$0.06\,dex.

In order to obtain separate constraints on $T_\mathrm{eff}$ and $\log g$ for the two components, we consider other, 
non-spectroscopic observations of the stars. We adopt the metallicity measurement of [Fe/H]=\,$-$0.30$\pm$0.06\,dex by \citet{Bruntt12}
for both components. {We expect both stars would have had the same initial composition because they would have formed at the same time from the same gas cloud. Effects such as gravitational settling could have subsequently led to different surface metallicities; however, because the stars have a similar mass, we consider these effects to be negligible.}

\subsection{Deriving individual $T_\mathrm{eff}$ and $\log g$}

\begin{table}
 \centering
   \caption{Stellar properties of HD\,176465\,A~and~B}
   \label{tbl-1}
  \begin{tabular}{lr@{$\pm$}lr@{$\pm$}l}
  \toprule
Property &\multicolumn{2}{c}{HD\,176465\,A} & \multicolumn{2}{c}{HD\,176465\,B}\\
\midrule
$Hp$           & 8.605 & 0.006 & 8.797 & 0.007 \\
$B_T$          & 9.193 & 0.018 & 9.394 & 0.019 \\
$V_T$          & 8.537 & 0.012 & 8.674 & 0.014 \\
Parallax (mas) & \multicolumn{4}{c}{20.18$\pm$0.74\enskip}   \\
Distance (pc)  & \multicolumn{4}{c}{49.6$\pm$1.8\enskip}   \\
Radial velocity (km\,s$^{-1}$) & \multicolumn{4}{c}{$-$30.5$\pm$0.4\enspace\quad} \\
$[\mathrm{Fe/H}]$ (dex) & \multicolumn{4}{c}{$-$0.30$\pm$0.06\quad} \\
$\nu_\mathrm{max}$ ($\mu$Hz)& 3260 & 30     & 3520 & 40 \\
$\Delta\nu$ ($\mu$Hz) & 146.79 & 0.12   & 155.42 & 0.13 \\
$\log (g/\mathrm{cm\,s^{-2}})$ & 4.463 & 0.005    & 4.492 & 0.006 \\
$T_\mathrm{eff}$ (K) & 5830 & 90      & 5740 & 90 \\
\bottomrule
\end{tabular}
\end{table}

We have separate photometric and asteroseismic measurements of each component with which we can obtain
separate constraints for each star. Separate magnitudes for each star are provided in the Hipparcos \citep{Perryman97} and Tycho-2 \citep{Hoeg00} catalogues, and are listed in Table \ref{tbl-1}.

From the apparent magnitudes and revised Hipparcos parallax \citep{VanLeeuwen07}, the absolute magnitude can be determined. Using a bolometric correction, the absolute magnitude can be used to find the stellar luminosity. Since
$L\propto R^2 T_\mathrm{eff}^4$, we only require the radius of the star to determine the effective temperature. This final
ingredient is provided by the asteroseismic scaling relation for radius,
\begin{equation}
\frac{R}{\mathrm{R}_\odot}\approx\left(\frac{\nu_\mathrm{max}}{\nu_\mathrm{max,\odot}}\right)\left(\frac{\Delta\nu}{\Delta\nu_\odot}\right)^{-2}\left(\frac{T_\mathrm{eff}}{\mathrm{T_{eff,\odot}}}\right)^{1/2}.\label{scale_radius}
\end{equation}
Combined with the luminosity, we then have,
\begin{equation}
\frac{T_\mathrm{eff}}{\mathrm{T_{eff,\odot}}}\approx\left(\frac{L}{\mathrm{L_\odot}}\right)^{1/5}\left(\frac{\nu_\mathrm{max}}{\nu_\mathrm{max,\odot}}\right)^{-2/5}\left(\frac{\Delta\nu}{\Delta\nu_\odot}\right)^{4/5}.\label{scale_teff}
\end{equation}

Bolometric corrections have been computed in the Hipparcos bandpass \citep{Bessell00} using the Castelli models discussed by \citet{Bessell98}.
The $Hp$-band bolometric corrections were determined by interpolating the model grid to the measured value of $B_T-V_T$, and the assumed
spectroscopic values of [Fe/H] and $\log g$. This bolometric correction was then applied to the absolute $Hp$-band
magnitude to determine the bolometric magnitude of the stars, and thus the luminosity.

Values of the asteroseismic parameters were measured from the {\it Kepler} data. Usual methods of determining $\nu_\mathrm{max}$,
such as heavily smoothing the power spectrum and fitting a Gaussian function, could not be applied in this case. This
is due to the frequencies of both stars existing within the same range. Instead we have used the results of the
peak-bagging to separate the frequencies from each star and fitted a Gaussian function to the mode amplitudes. The
large separations, $\Delta\nu$, were derived from a linear least-squares fit to the $l$=0 frequencies. The measured
asteroseismic parameters for each star are listed in Table \ref{tbl-1}.

Equation (\ref{scale_teff}) was then used to provide a first estimate of $T_\mathrm{eff}$. The bolometric correction
was based on the value of $\log g$ from spectroscopy. The measurement of $\log g$ for each star was then improved 
using the asteroseismic scaling relation for the frequency of maximum power \citep{Brown91,Kjeldsen95,Belkacem11}, 
\begin{equation}
\nu_\mathrm{max} \propto \nu_\mathrm{ac} \propto \frac{c}{H} \propto g T_\mathrm{eff}^{-1/2}, \label{scale_numax}
\end{equation} 
where $\nu_\mathrm{ac}$ is the acoustic cut-off frequency, $c$ is the sound speed, 
$H=-\left(\mathrm{d}\ln\rho/\mathrm{d}r\right)^{-1}$ is the density scale height. We iterated several times, redetermining the bolometric
correction for the new values of $\log g$ to find updated values of $T_\mathrm{eff}$ until the values became stable.

Uncertainties in $T_\mathrm{eff}$ and $\log g$ were determined through Monte Carlo simulations, assuming Gaussian uncertainties
in all input parameters. The derived quantities are included in Table \ref{tbl-1}, and are consistent with the spectroscopic values determined for the combined system.

%______________________________________________________________

\section{Modelling}\label{sec:model}

We have used four different modelling approaches to determine the properties of HD\,176465\,A~and~B using the measured oscillation frequencies determined in Sect.~\ref{sec:mode_param} and the non-seismic parameters determined in Sect.~\ref{sec:non_seism}. Following \citet{Davies14}, we corrected the frequencies for the line-of-sight Doppler velocity of the system \citep[$-$30.5$\pm$0.4\,km\,s$^{-1}$;][]{Gontcharov06}. Although the relative Doppler velocities of the two components is unknown, it cannot be larger than the value of $v \sin i$ determined from their combined spectra when analysed as a single star \citep[3.0\,km\,s$^{-1}$;][]{Bruntt12}, which we have taken into account when determining the uncertainties of the corrected frequencies. The radial-velocity-corrected frequencies are provided alongside the other mode properties in Tables~\ref{tab:Luke_freq}~and~\ref{tab:Leia_freq}.

Since each modelling method varies in input physics, evolutionary and pulsation code, and minimization technique, the variety of methods we employed provides an indication of the systematic uncertainty in the determined parameters. The four methods were the Asteroseismic Modeling Portal (AMP), the \textsc{astec} Fitting (ASTFIT) method, the Bayesian Stellar Algorithm (BASTA), and modelling with the Modules for Experiments in Stellar Astrophysics (\textsc{mesa}) code.

The AMP, originally introduced by \citet{Metcalfe09}, uses a parallel genetic algorithm \citep{Metcalfe03} to optimize the match between a set of observations and models produced by the Aarhus stellar evolution and pulsation codes \citep[\textsc{astec} and \textsc{adipls};][]{C-D08a,C-D08b}. The AMP code has been in continuous development \citep{Mathur12,Metcalfe14,Metcalfe15}, and here we use the most recent version (Creevey et al., submitted to A\&A). The models use the \citet{GrevesseSauval98} solar mixture, with the OPAL 2005 equation of state \citep{Rogers02}, OPAL opacities at high temperatures \citep{Iglesias96} supplemented with those of \citet{Ferguson05} at low temperatures, and NACRE nuclear reaction rates \citep{Angulo99,Angulo05}.

The ASTFIT and BASTA methods are both described by \citet{SilvaAguirre15}. While both methods use the same input physics (including a solar-calibrated mixing length parameter, $\alpha$), the ASTFIT method uses models calculated with \textsc{astec}, whereas the BASTA uses models calculated with the Garching Stellar Evolution Code \citep[\textsc{garstec};][]{Weiss08}. Furthermore, they differ in the asteroseismic variables that are fitted. The ASTFIT method matches individual frequencies, with surface effects corrected via a scaled function determined by frequency corrections found in the solar case, whereas the BASTA fits frequency ratios, which are relatively insensitive to surface effects \citep{Roxburgh03}. Both ASTFIT and BASTA have assumed a linear helium-to-metal enrichment law, $\Delta Y/\Delta Z\,=\,1.4$ on the basis of big bang nucleosythesis primordial values \citep[$Z_0\,=\,0.0$, $Y_0\,=\,0.248$;][]{Steigman10} and initial abundances from calibration of solar models. ASTFIT adopts the \citet{GrevesseNoels93} solar mixture, while BASTA uses the mixture of \citet{GrevesseSauval98}.

\begin{table*}
 \centering
  \caption{Modelling results for HD\,176465\,A}
  \label{tbl:modelsA}
  \begin{tabular}{lccccc}
  \toprule
Property & AMP & ASTFIT & BASTA & MESA1 & MESA2 \\
\midrule
Mass (M$_\odot$)   &  $0.930\pm0.04$   &  $0.952\pm0.015$  & $0.960^{+0.010}_{-0.011}$ &   $0.95\pm0.03$   &  $0.99\pm0.02 $  \\ 
Radius (R$_\odot$) &  $0.918\pm0.015$  &  $0.927\pm0.005$  & $0.928^{+0.006}_{-0.003}$ &  $0.926\pm0.011$  & $0.939\pm0.006$ \\ 
Age (Gyr)          &    $3.0\pm0.4$    &    $3.2\pm0.5$    &        $2.8\pm0.3$        &    $3.2\pm0.2$    &  $3.01\pm0.12$  \\ 
$Z_i$              & $0.0085\pm0.0010$ & $0.0103\pm0.0010$ &      $0.011\pm0.004$      & $0.0102\pm0.0009$ & $0.0094\pm0.0009$\\ 
$Y_i$              &  $0.258\pm0.024$  &  $0.262\pm0.003$  &      $0.265\pm0.002$      &   $0.25\pm0.02$   &  $0.23\pm0.02$  \\ 
$\alpha$           &   $1.90\pm0.18$   &       1.80        &           1.791           &   $1.57\pm0.11$   &      1.79       \\ 
\bottomrule
\end{tabular}
\end{table*}

\begin{table*}
 \centering
  \caption{Modelling results for HD\,176465\,B}
  \label{tbl:modelsB}
  \begin{tabular}{lccccc}
  \toprule
Property & AMP & ASTFIT & BASTA & MESA1 & MESA2 \\
\midrule
Mass (M$_\odot$)   &  $0.930\pm0.02$   &   $0.92\pm0.02$   & $0.929^{+0.010}_{-0.011}$ &   $1.02\pm0.07$   &  $0.97\pm0.04$  \\ 
Radius (R$_\odot$) &  $0.885\pm0.006$  &  $0.883\pm0.007$  & $0.886^{+0.003}_{-0.006}$ &  $0.919\pm0.021$  & $0.899\pm0.013$ \\ 
Age (Gyr)          &    $2.9\pm0.5$    &    $3.4\pm0.9$    &        $3.2\pm0.4$        &    $2.9\pm0.4$    &   $3.18\pm0.31$   \\ 
$Z_i$              & $0.0085\pm0.0007$ & $0.0096\pm0.0011$ &      $0.011\pm0.004$      & $0.0124\pm0.0015$ & $0.0122\pm0.0011$\\ 
$Y_i$              &  $0.246\pm0.013$  &  $0.262\pm0.003$  &      $0.265\pm0.002$      &   $0.21\pm0.04$   &  $0.24\pm0.04$  \\ 
$\alpha$           &   $1.94\pm0.12$   &       1.80        &           1.791           &    $2.05\pm0.28$   &     1.79        \\ 
\bottomrule
\end{tabular}
\end{table*}

Finally, the system was modelled using the \textsc{mesa}\footnote{\url{http://mesa.sourceforge.net}} \citep[revision 7624;][]{Paxton11, Paxton13, Paxton15}. These stellar models used the solar mixture of \citet{GrevesseSauval98} with opacities from the OPAL tables \citep{Iglesias96} at high temperatures and from \citet{Ferguson05} at low temperatures, and the atmospheric structure was integrated from the photosphere to an optical depth $\tau=10^{-4}$ using the standard grey Eddington atmosphere. Convection was described using the standard mixing-length theory \citep[e.g.][]{BoehmVitense58} without any convective overshooting or semiconvection. Gravitational settling was included according to the method of \citet{Thoul94}. Nuclear reactions rates were drawn from tables provided by the NACRE collaboration \citep{Angulo99} or, when those are unavailable, the tables by \citet{Caughlan88}. We also used newer rates for the specific reactions ${}^{14}\mathrm{N}(p,\gamma){}^{15}\mathrm{O}$ \citep{Imbriani05} and ${}^{12}\mathrm{C}(\alpha,\gamma){}^{16}\mathrm{O}$ \citep{Kunz02}. Oscillation mode frequencies were computed by internal calls to \textsc{adipls}. Model frequencies were corrected for surface effects using the two-term (or \emph{combined}) correction described by \citet{Ball14}.

For a given choice of mass $M$, initial metal abundance $Z_i$, initial helium abundance $Y_i$ and mixing-length parameter $\alpha$, \textsc{mesa} evolved a stellar model from a pre-main-sequence model with central temperature $T_c=9\times10^{6}\,\mathrm{K}$ and found the best-fitting age and surface correction parameters for those input parameters. Starting from an initial sample of 11 guesses from grid-based modelling, the input parameters were optimized with the Nelder--Mead algorithm \citep{Nelder65}, accelerated with linear extrapolations of the observables as functions of the initial parameters. When further iterations failed to produce better-fitting parameters, new parameter guesses were randomly drawn uniformly from within the $1\sigma$ uncertainty region and these used to continue the Nelder--Mead simplex if the random guesses were better than any element of the simplex. Once the results appeared to have converged, a final simplex was run using the best-fitting model as the initial guess to ensure that it represented a true (local) minimum. Uncertainties were determined by selecting the model with smallest $\chi^2$, which we denote $\chi^2_0$, shrinking points with $\chi^2>\chi^2_0+1$ towards the best-fitting model by rescaling their distances by $\sqrt{\chi^2-\chi^2_0}$, and finding the minimum bounding ellipsoid around the rescaled sample.

The modelling results are presented in Tables~\ref{tbl:modelsA}~and~\ref{tbl:modelsB} for HD\,176465\,A~and~B, respectively. The various methods agree rather well for both stars. All methods converged on models for HD\,176465\,A that have a mass of 0.95$\pm$0.02\,M$_\odot$ and a radius of 0.93$\pm$0.01\,R$_\odot$. For HD\,176465\,B, AMP, ASTFIT and BASTA all found a mass of 0.93$\pm$0.01\,M$_\odot$ and a radius of 0.89$\pm$0.01\,R$_\odot$. 

The best-fitting \textsc{mesa} model of HD\,176465\,B has a higher mass than the model of HD\,176465\,A, and an initial helium mass fraction $Y_\mathrm{i}$ significantly lower than the primordial value from standard Big Bang nucleosynthesis \citep[$Y\,=\,0.2482\pm0.0007$;][]{Steigman10}. A further difference between these MESA models can be noted in the mixing-length parameter, with the values for the HD\,176465\,A~and~B models being lower and higher, respectively, than the solar-calibrated value. Calibration of the mixing length against 3D models of convection suggests that neither of these behaviours is to be expected for stars so similar to the Sun \citep{Trampedach14}.

\. {A further set of \textsc{MESA} models run with the mixing-length parameter kept fixed at the solar-calibrated value (designated MESA2 in Tables~\ref{tbl:modelsA}~and~\ref{tbl:modelsB}) results in a consistent initial helium fraction for both stars and a higher mass for the primary component, as expected. However, the initial helium fraction of the best-fitting MESA2 models remain below the primordial value, which may partly explain why the masses and radii of these models are larger than those obtained with AMP, ASTFIT and BASTA.} Similarly, the best-fitting AMP model of HD\,176465\,B has a comparable mass to the AMP model of HD\,176465\,A while also having a low initial helium abundance. This degeneracy between mass and helium abundance is well-known in asteroseismic modelling of solar-like oscillations \citep[see][]{SilvaAguirre15}. As discussed, the ASTFIT and BASTA methods have circumvented this problem by enforcing $\Delta Y/\Delta Z\,=\,1.4$.

Significantly, although the ages of both components were not {\it a priori} constrained to be the same, all models agreed on the same age of both stars, 3.0$\pm$0.5\,Gyr. 

%______________________________________________________________

\section{Stellar Activity}\label{sec:activity}

\begin{figure}
\centering
\resizebox{\hsize}{!}{\includegraphics{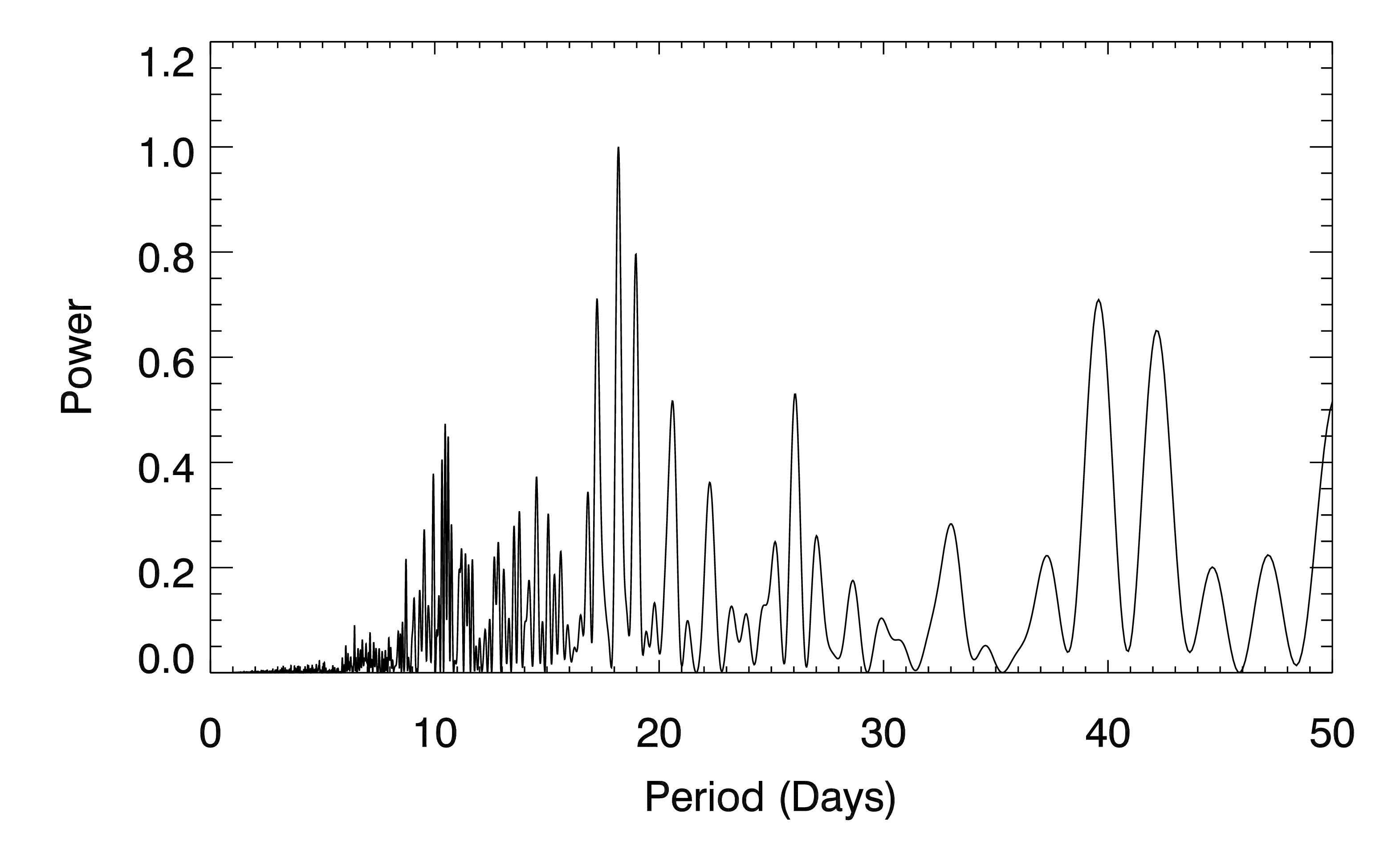}}
\caption{Periodogram of the long cadence data showing evidence of rotational modulation at a period of $\sim$18\,days.}
\label{fig:rot}
\end{figure}

In addition to solar-like oscillations, the time series of HD\,176465 also shows variation on longer time scales, indicative of spot-induced rotational modulation. The periodogram, shown in Fig.~\ref{fig:rot}, has a clear signal with a period of $\sim18\,$d. 

The light curve was analysed by several teams using a variety of methods, which have been compared by \citet{Aigrain15}. \citet{Garcia14} reported a period value from the Q0--14 time series of 17.6$\pm$2.3\,d. \citet{McQuillan14} found a period of 19.2$\pm$0.8\,d, although this period did not meet their significance criterion, which can be attributed to the rotational modulation not being particularly stable. This is probably due to the spots being small or short-lived, but it is possible that beating between two closely-spaced frequencies, corresponding to the rotation periods of each star, may also contribute. The periods measured by \citet{Garcia14} and \citet{McQuillan14} are consistent with the values determined from the rotational splitting of non-radial modes in Sect.~\ref{sec:mode_param} for both stars.

Since the rotation rates of stars are known to decrease as they age, the rotation period may be used to provide an estimate of the age using suitably calibrated gyrochronology relations \citep{Skumanich72,Barnes10,Epstein14}. Using the gyrochronology relation of \citet{Barnes10}, and using $P_\mathrm{rot}\,=\,18\,\pm2$\,d, we find ages of $2.5^{+0.6}_{-0.5}$~ and~$2.1^{+0.5}_{-0.4}$\,Gyr for the A and B components, respectively. \citet{Garcia14} calibrated a simpler gyrochronology relation between age and period for a sample of stars with asteroseismic ages; using this relation we find an age of $3.2^{+1.2}_{-0.8}$\,Gyr.

These gyrochronological ages are in good agreement with the asteroseismic ages obtained in Sect.~\ref{sec:model}. This is in contrast to the recent study by \citet{Angus15}, who found a significant disagreement between asteroseismic and gyrochronological ages in {\it Kepler} field stars. However, \citet{VanSaders16} have explained that disagreement as being a result of weakened magnetic braking in old stars. Since the HD\,176465 system is still relatively young, the standard gyrochronology relations still appear to hold.

\citet{Garcia14} also investigated the level of photospheric magnetic activity in HD\,176465. They computed the standard deviation of the time series, which has been used as a photometric magnetic activity index, denoted $S_\mathrm{ph}$ \citep{Garcia10,Mathur14}. \citet{Garcia14} found $S_\mathrm{ph}$\,=\,248$\,\pm$\,4\,ppm, which is close to the value for the Sun during its maximum activity of Cycle 23, as measured from VIRGO/SPM \citep[see][]{Mathur14}. Furthermore, the $S_\mathrm{ph}$ obtained is a lower limit since the inclination angle is $\sim50^\circ$. Although it is difficult to distinguish between the contributions of each star, it is likely that this measurement represents a higher photospheric magnetic activity than the Sun during the last activity maximum.

Additionally, HD\,176465 was amongst a sample of {\it Kepler} targets monitored for excess flux in the Ca~H~and~K lines with the Fibre-fed Echelle Spectrograph (FIES) on the Nordic Optical Telescope by \citet{Karoff13}, although it was dropped from their analysis due to its binary nature. Excess flux in the cores of these lines arises from magnetic sources \citep[see][]{Schrijver89}. Analysing these FIES spectra following the methods of \citet{Karoff13}, we find the excess flux to be $\Delta\mathcal{F}_\mathrm{Ca}$\,=\,6.02\,$\pm$\,0.28, from which we derive the Mount Wilson $S$ index, 0.170\,$\pm$\,0.003. As for the photometric measurement of activity, both of these values reflect a relatively high activity level.

%__________________________________________________________

\section{Conclusions}

Situated at only $49.6 \pm 1.8$ pc, the HD\,176465 binary system is sufficiently bright to reveal solar-like oscillations in both components. Only four other systems, namely $\alpha$\,Cen, 16\,Cyg, HD\,177412, and HD\,176071, are known to share this property. With both stars having solar-like oscillations with $\nu_\mathrm{max}\approx3500\,\mu$Hz and $\Delta\nu\,\approx\,150\,\mu$Hz, and with an effective temperature of $\approx\,5700$\,K, the system is comprised of two similar solar analogues.

Detailed modelling of the power spectrum has enabled us to extract the frequencies, linewidths and heights of 45 modes for HD\,176465\,A, and 25 modes for HD\,176465\,B, up to an angular degree of $l=2$ and with a statistical significance greater than $75\%$. 

The rotational splittings of the non-radial modes reveal both stars to have a similar rotation period of $\sim18\,$d and inclination of $\sim50^\circ$. This rotation period is in agreement with that determined from the rotational modulation caused by the presence of starspots. Gyrochronology relations imply an age of $\sim3\,$Gyr.

Precise estimates for the fundamental stellar parameters have been derived through detailed modelling of both stars. The four methods used to model these stars agree on the stellar properties in almost all cases. The masses of the stars are $M_A\,=\,0.95\pm0.02\,$M$_\odot$ and $M_B\,=\,0.93\pm0.01\,$M$_\odot$, and their radii are $R_A\,=\,0.93\pm0.01\,$R$_\odot$ and $R_B\,=\,0.89\pm0.01\,$R$_\odot$. From the modelling, the ages of both stars are found to be $3.0\pm0.5$\,Gyrs, in agreement with the gyrochronological value, and consistent with a synchronous formation.

While main-sequence systems such as this are rare, they provide important tests of stellar structure and evolution, and of asteroseismology. \citet{Miglio14} predicted there should be more such systems amongst the many red giants observed by {\it Kepler}. The absence in the literature of systems with clear oscillations in two red giants is possibly a consequence of the difficulty in recognising the typical pattern of solar-like oscillations when modes overlap in frequency. While the overlapping modes of HD\,176465 have presented a (thus far) unique challenge in determining mode parameters, overlapping modes of red giants, with their many mixed $l=1$ modes, will present a far greater challenge. Despite these challenges, such systems will provide further tests of stellar structure and evolution when they are found, just as systems with a single oscillating red giant have already \citep[e.g.][]{Hekker10b,Frandsen13,Gaulme13,Gaulme14,Beck14,Rawls16}. With the future missions {\it TESS} \citep{Ricker15,Campante16} and {\it PLATO} \citep{Rauer14} to provide asteroseismic data for many more systems, there will be good opportunities to find more such asteroseismic binary systems.

%______________________________________________________________

\begin{acknowledgements}
The authors gratefully acknowledge the {\it Kepler} Science Team and all those who have contributed to the {\it Kepler Mission} 
for their tireless efforts which have made these results possible. Funding for the {\it Kepler Mission} was provided by 
NASA’s Science Mission Directorate. 
Funding for the Stellar Astrophysics Centre is provided by The Danish National Research Foundation (Grant agreement no.: DNRF106). The research was supported by the ASTERISK project (ASTERoseismic Investigations with SONG and {\it Kepler}) funded by the European Research Council (Grant agreement no.: 267864). TRW and VSA acknowledge support from the Villum Foundation (research grant 10118) and the Instrument Centre for Danish Astrophysics. CK also acknowledges the support of the Villum Foundation.
OB, LG and MB acknowledge support by the Center for Space Science at the NYU Abu Dhabi Institute under grant G1502.
WHB acknowledges research funding by Deutsche Forschungsgemeinschaft (DFG) under grant SFB 963/1 ``Astrophysical flow instabilities and turbulence” (Project A18).
GRD, WJC, TLC, and YE acknowledge the support of the UK Science and Technology Facilities Council (STFC).
RAG, BM, and D.~Salabert acknowledges funding from the CNES and the ANR (Agence Nationale de la Recherche, France) program IDEE (n ANR-12-BS05-0008) ``Interaction Des \'Etoiles et des Exoplan\`etes''.
The research leading to the presented results has received funding from the
European Research Council under the European Community's Seventh Framework
Programme (FP7/2007-2013) / ERC grant agreement no 338251 (StellarAges).
R.~Howe acknowledges computing support from the National Solar Observatory.
DH acknowledges support by the Australian Research Council's Discovery Projects funding scheme (project number DE140101364) and support by the National Aeronautics and Space Administration under Grant NNX14AB92G issued through the Kepler Participating Scientist Program.
SM acknowledges support from the NASA grant NNX12AE17G.
This research made use of NASA's Astrophysics Data System and the SIMBAD database, operated at CDS, Strasbourg, France.
TSM acknowledges NASA grants NNX13AE91G and NNX15AF13G. Computational
time at the Texas Advanced Computing Center was provided through XSEDE
allocation TG-AST090107.
\end{acknowledgements}

%-------------------------------------------------------------------

\bibliographystyle{aa}%Used BibTeX style is unsrt
\bibliography{references}

\begin{table*}
\centering
\caption{Fitted mode parameters of HD\,176465\,A}
\label{tab:Luke_freq}
\begin{tabular}{cccccccc}
\toprule
$n$ & $\ell$ & Frequency  & RV Corr. Freq. & Height          & Linewidth & Amplitude  & $P(M_{mode} | \boldsymbol{y})$ \\ 
    &        & ($\mu$Hz)  & ($\mu$Hz)      & (ppm$^2/\mu$Hz) & ($\mu$Hz) & (ppm)      & \% \\
\midrule
12 & 2 & 2121.39$\pm$1.67 & 2121.18$\pm$1.67 & $0.023^{+0.010}_{-0.004}$ & $1.16^{+0.32}_{-0.58}$ & $0.20^{+0.08}_{-0.06}$ & $84.0$ \\ \midrule
13 & 0 & 2132.67$\pm$0.61 & 2132.45$\pm$0.61 & $0.043^{+0.018}_{-0.006}$ & $1.12^{+0.35}_{-0.52}$ & $0.27^{+0.11}_{-0.08}$ & $99.8$ \\
   & 1 & 2202.44$\pm$0.39 & 2202.22$\pm$0.39 & $0.065^{+0.027}_{-0.009}$ & $0.86^{+0.61}_{-0.15}$ & $0.29^{+0.17}_{-0.04}$ & $80.3$ \\
   & 2 & 2268.74$\pm$1.17 & 2268.51$\pm$1.17 & $0.029^{+0.007}_{-0.006}$ & $0.89^{+0.56}_{-0.35}$ & $0.21^{+0.04}_{-0.05}$ & $79.5$ \\ \midrule
14 & 0 & 2280.58$\pm$0.90 & 2280.34$\pm$0.90 & $0.054^{+0.013}_{-0.010}$ & $0.92^{+0.53}_{-0.41}$ & $0.29^{+0.02}_{-0.09}$ & $96.2$ \\
   & 1 & 2348.98$\pm$0.24 & 2348.74$\pm$0.24 & $0.082^{+0.019}_{-0.015}$ & $1.39^{+0.47}_{-0.27}$ & $0.41^{+0.14}_{-0.06}$ & $83.4$ \\
   & 2 & 2415.63$\pm$1.97 & 2415.38$\pm$1.97 & $0.047^{+0.012}_{-0.012}$ & $2.12^{+0.79}_{-1.28}$ & $0.37^{+0.05}_{-0.10}$ & $18.4$ \\ \midrule
15 & 0 & 2428.06$\pm$0.38 & 2427.81$\pm$0.38 & $0.087^{+0.022}_{-0.021}$ & $2.25^{+0.86}_{-1.47}$ & $0.52^{+0.08}_{-0.16}$ & $96.4$ \\
   & 1 & 2496.73$\pm$0.33 & 2496.48$\pm$0.33 & $0.131^{+0.034}_{-0.032}$ & $2.03^{+0.31}_{-1.21}$ & $0.59^{+0.08}_{-0.13}$ &  $100$ \\
   & 2 & 2561.06$\pm$1.65 & 2560.80$\pm$1.65 & $0.055^{+0.012}_{-0.009}$ & $1.43^{+0.56}_{-0.62}$ & $0.34^{+0.05}_{-0.05}$ & $39.3$ \\ \midrule
16 & 0 & 2575.83$\pm$0.83 & 2575.57$\pm$0.83 & $0.102^{+0.021}_{-0.017}$ & $1.33^{+0.65}_{-0.55}$ & $0.46^{+0.07}_{-0.07}$ & $84.2$ \\
   & 1 & 2642.66$\pm$0.10 & 2642.39$\pm$0.10 & $0.154^{+0.032}_{-0.025}$ & $1.68^{+0.27}_{-0.59}$ & $0.62^{+0.04}_{-0.05}$ &  $100$ \\
   & 2 & 2706.15$\pm$0.74 & 2705.87$\pm$0.74 & $0.074^{+0.014}_{-0.013}$ & $1.87^{+0.37}_{-0.49}$ & $0.45^{+0.05}_{-0.04}$ & $98.0$ \\ \midrule
17 & 0 & 2720.12$\pm$0.28 & 2719.84$\pm$0.28 & $0.138^{+0.026}_{-0.024}$ & $1.92^{+0.40}_{-0.50}$ & $0.63^{+0.06}_{-0.05}$ &  $100$ \\
   & 1 & 2788.08$\pm$0.36 & 2787.80$\pm$0.36 & $0.208^{+0.038}_{-0.036}$ & $1.56^{+0.23}_{-0.28}$ & $0.70^{+0.05}_{-0.04}$ &  $100$ \\
   & 2 & 2852.75$\pm$0.38 & 2852.46$\pm$0.38 & $0.202^{+0.024}_{-0.021}$ & $1.17^{+0.18}_{-0.17}$ & $0.60^{+0.04}_{-0.03}$ & $99.6$ \\ \midrule
18 & 0 & 2865.49$\pm$0.09 & 2865.19$\pm$0.10 & $0.381^{+0.046}_{-0.039}$ & $1.11^{+0.19}_{-0.20}$ & $0.81^{+0.05}_{-0.04}$ &  $100$ \\
   & 1 & 2934.23$\pm$0.12 & 2933.93$\pm$0.12 & $0.571^{+0.068}_{-0.059}$ & $1.09^{+0.11}_{-0.18}$ & $0.98^{+0.04}_{-0.04}$ &  $100$ \\
   & 2 & 2999.34$\pm$0.22 & 2999.03$\pm$0.22 & $0.275^{+0.044}_{-0.034}$ & $1.02^{+0.19}_{-0.13}$ & $0.67^{+0.03}_{-0.03}$ &  $100$ \\ \midrule
19 & 0 & 3012.10$\pm$0.11 & 3011.79$\pm$0.11 & $0.519^{+0.082}_{-0.065}$ & $1.01^{+0.20}_{-0.13}$ & $0.91^{+0.05}_{-0.05}$ &  $100$ \\
   & 1 & 3081.19$\pm$0.10 & 3080.88$\pm$0.11 & $0.778^{+0.122}_{-0.097}$ & $1.11^{+0.12}_{-0.09}$ & $1.17^{+0.05}_{-0.04}$ &  $100$ \\
   & 2 & 3146.38$\pm$0.21 & 3146.06$\pm$0.21 & $0.327^{+0.025}_{-0.027}$ & $1.20^{+0.11}_{-0.11}$ & $0.78^{+0.03}_{-0.03}$ &  $100$ \\ \midrule
20 & 0 & 3158.74$\pm$0.02 & 3158.41$\pm$0.04 & $0.620^{+0.047}_{-0.050}$ & $1.21^{+0.13}_{-0.11}$ & $1.08^{+0.04}_{-0.04}$ &  $100$ \\
   & 1 & 3227.97$\pm$0.09 & 3227.64$\pm$0.09 & $0.924^{+0.070}_{-0.074}$ & $1.34^{+0.09}_{-0.10}$ & $1.39^{+0.05}_{-0.05}$ &  $100$ \\
   & 2 & 3293.79$\pm$0.17 & 3293.45$\pm$0.17 & $0.333^{+0.026}_{-0.031}$ & $1.45^{+0.15}_{-0.13}$ & $0.87^{+0.03}_{-0.03}$ &  $100$ \\ \midrule
21 & 0 & 3305.62$\pm$0.08 & 3305.28$\pm$0.08 & $0.629^{+0.049}_{-0.058}$ & $1.47^{+0.16}_{-0.14}$ & $1.20^{+0.04}_{-0.04}$ &  $100$ \\
   & 1 & 3375.32$\pm$0.10 & 3374.97$\pm$0.10 & $0.944^{+0.073}_{-0.086}$ & $1.82^{+0.12}_{-0.13}$ & $1.63^{+0.05}_{-0.05}$ &  $100$ \\
   & 2 & 3441.16$\pm$0.36 & 3440.81$\pm$0.35 & $0.242^{+0.017}_{-0.015}$ & $2.12^{+0.16}_{-0.16}$ & $0.90^{+0.03}_{-0.03}$ &  $100$ \\ \midrule
22 & 0 & 3452.35$\pm$0.11 & 3452.00$\pm$0.11 & $0.457^{+0.031}_{-0.028}$ & $2.17^{+0.17}_{-0.17}$ & $1.25^{+0.04}_{-0.05}$ &  $100$ \\
   & 1 & 3522.07$\pm$0.14 & 3521.72$\pm$0.14 & $0.686^{+0.047}_{-0.042}$ & $2.22^{+0.11}_{-0.14}$ & $1.54^{+0.05}_{-0.04}$ &  $100$ \\
   & 2 & 3588.47$\pm$0.26 & 3588.10$\pm$0.26 & $0.175^{+0.016}_{-0.014}$ & $2.24^{+0.22}_{-0.19}$ & $0.79^{+0.03}_{-0.03}$ &  $100$ \\ \midrule
23 & 0 & 3599.69$\pm$0.13 & 3599.32$\pm$0.14 & $0.330^{+0.030}_{-0.025}$ & $2.24^{+0.25}_{-0.21}$ & $1.08^{+0.04}_{-0.04}$ &  $100$ \\
   & 1 & 3669.46$\pm$0.16 & 3669.08$\pm$0.16 & $0.495^{+0.045}_{-0.037}$ & $2.23^{+0.21}_{-0.19}$ & $1.32^{+0.06}_{-0.06}$ &  $100$ \\
   & 2 & 3735.42$\pm$0.63 & 3735.04$\pm$0.63 & $0.129^{+0.018}_{-0.014}$ & $2.21^{+0.33}_{-0.33}$ & $0.67^{+0.03}_{-0.04}$ &  $100$ \\ \midrule
24 & 0 & 3746.85$\pm$0.16 & 3746.47$\pm$0.16 & $0.243^{+0.033}_{-0.025}$ & $2.21^{+0.35}_{-0.36}$ & $0.92^{+0.05}_{-0.05}$ &  $100$ \\
   & 1 & 3817.25$\pm$0.24 & 3816.86$\pm$0.24 & $0.365^{+0.049}_{-0.038}$ & $2.88^{+0.48}_{-0.36}$ & $1.30^{+0.06}_{-0.07}$ &  $100$ \\
   & 2 & 3884.21$\pm$1.72 & 3883.81$\pm$1.72 & $0.073^{+0.011}_{-0.009}$ & $3.54^{+0.71}_{-0.51}$ & $0.64^{+0.04}_{-0.05}$ & $99.7$ \\ \midrule
25 & 0 & 3895.17$\pm$0.18 & 3894.77$\pm$0.18 & $0.137^{+0.021}_{-0.017}$ & $3.64^{+0.76}_{-0.55}$ & $0.90^{+0.05}_{-0.07}$ &  $100$ \\
   & 1 & 3965.06$\pm$0.40 & 3964.65$\pm$0.40 & $0.206^{+0.031}_{-0.026}$ & $4.78^{+0.59}_{-0.60}$ & $1.24^{+0.10}_{-0.09}$ &  $100$ \\
   & 2 & 4033.13$\pm$1.44 & 4032.72$\pm$1.44 & $0.047^{+0.008}_{-0.005}$ & $5.67^{+1.40}_{-0.86}$ & $0.65^{+0.07}_{-0.05}$ & $99.3$ \\ \midrule
26 & 0 & 4043.26$\pm$0.02 & 4042.85$\pm$0.05 & $0.088^{+0.015}_{-0.010}$ & $5.80^{+1.49}_{-0.90}$ & $0.90^{+0.10}_{-0.08}$ &  $100$ \\
   & 1 & 4114.50$\pm$0.50 & 4114.08$\pm$0.50 & $0.131^{+0.022}_{-0.014}$ & $5.26^{+0.87}_{-0.77}$ & $1.06^{+0.05}_{-0.07}$ &  $100$ \\
   & 2 & 4181.86$\pm$0.89 & 4181.43$\pm$0.89 & $0.023^{+0.008}_{-0.006}$ & $4.59^{+1.60}_{-1.35}$ & $0.40^{+0.03}_{-0.04}$ & $81.2$ \\ \midrule
27 & 0 & 4192.29$\pm$0.46 & 4191.86$\pm$0.46 & $0.044^{+0.015}_{-0.012}$ & $4.47^{+1.79}_{-1.38}$ & $0.55^{+0.05}_{-0.06}$ & $98.8$ \\
   & 1 & 4263.57$\pm$0.17 & 4263.14$\pm$0.18 & $0.066^{+0.022}_{-0.017}$ & $4.21^{+2.32}_{-1.36}$ & $0.67^{+0.06}_{-0.08}$ & $98.3$ \\
   & 2 & 4329.12$\pm$2.46 & 4328.68$\pm$2.46 & $0.009^{+0.002}_{-0.002}$ & $4.39^{+2.43}_{-2.08}$ & $0.24^{+0.08}_{-0.08}$ & $89.7$ \\ \midrule
28 & 0 & 4341.82$\pm$0.91 & 4341.38$\pm$0.91 & $0.017^{+0.003}_{-0.004}$ & $4.40^{+2.49}_{-2.20}$ & $0.34^{+0.11}_{-0.11}$ & $98.4$ \\
   & 1 & 4411.38$\pm$0.27 & 4410.93$\pm$0.27 & $0.025^{+0.004}_{-0.005}$ & $4.52^{+2.69}_{-2.93}$ & $0.42^{+0.14}_{-0.19}$ & $44.9$ \\
\bottomrule
\end{tabular}
\end{table*}

\begin{table*}
\centering
\caption{Fitted mode parameters of HD\,176465\,B}
\label{tab:Leia_freq}
\begin{tabular}{cccccccc}
\toprule
$n$ & $\ell$ & Frequency  & RV Corr. Freq. & Height          & Linewidth & Amplitude  & $P(M_{mode} | \boldsymbol{y})$ \\ 
    &        & ($\mu$Hz)  & ($\mu$Hz)      & (ppm$^2/\mu$Hz) & ($\mu$Hz) & (ppm)      & \% \\ \midrule
15 & 2 & 2709.26$\pm$2.31 & 2708.98$\pm$2.31 & $0.033^{+0.011}_{-0.008}$ & $0.19^{+0.33}_{-0.10}$ & $0.11^{+0.05}_{-0.04}$ & $59.9$ \\ \midrule
16 & 0 & 2725.97$\pm$1.56 & 2725.69$\pm$1.56 & $0.061^{+0.019}_{-0.016}$ & $0.21^{+0.29}_{-0.08}$ & $0.16^{+0.06}_{-0.04}$ & $37.4$ \\
   & 1 & 2797.21$\pm$0.43 & 2796.92$\pm$0.43 & $0.092^{+0.029}_{-0.023}$ & $0.34^{+0.07}_{-0.15}$ & $0.20^{+0.07}_{-0.03}$ & $40.2$ \\
   & 2 & 2864.06$\pm$1.63 & 2863.77$\pm$1.63 & $0.087^{+0.017}_{-0.016}$ & $0.38^{+0.15}_{-0.20}$ & $0.22^{+0.04}_{-0.07}$ & $22.6$ \\ \midrule
17 & 0 & 2879.67$\pm$0.47 & 2879.38$\pm$0.47 & $0.164^{+0.032}_{-0.030}$ & $0.38^{+0.19}_{-0.21}$ & $0.30^{+0.07}_{-0.10}$ & $41.4$ \\
   & 1 & 2951.76$\pm$0.21 & 2951.46$\pm$0.21 & $0.247^{+0.048}_{-0.045}$ & $0.58^{+0.11}_{-0.09}$ & $0.48^{+0.05}_{-0.06}$ &  $100$ \\
   & 2 & 3018.83$\pm$1.24 & 3018.52$\pm$1.24 & $0.078^{+0.013}_{-0.013}$ & $0.78^{+0.09}_{-0.11}$ & $0.31^{+0.02}_{-0.03}$ & $15.9$ \\ \midrule
18 & 0 & 3032.77$\pm$0.42 & 3032.46$\pm$0.42 & $0.147^{+0.025}_{-0.025}$ & $0.83^{+0.09}_{-0.14}$ & $0.43^{+0.04}_{-0.04}$ & $75.7$ \\
   & 1 & 3105.54$\pm$0.21 & 3105.23$\pm$0.21 & $0.220^{+0.037}_{-0.037}$ & $1.38^{+0.17}_{-0.16}$ & $0.69^{+0.07}_{-0.06}$ &  $100$ \\
   & 2 & 3173.61$\pm$0.71 & 3173.29$\pm$0.71 & $0.086^{+0.015}_{-0.013}$ & $1.90^{+0.25}_{-0.21}$ & $0.51^{+0.03}_{-0.03}$ & $42.9$ \\ \midrule
19 & 0 & 3187.74$\pm$0.26 & 3187.41$\pm$0.26 & $0.162^{+0.029}_{-0.025}$ & $2.01^{+0.27}_{-0.23}$ & $0.72^{+0.05}_{-0.04}$ &  $100$ \\
   & 1 & 3260.58$\pm$0.15 & 3260.25$\pm$0.15 & $0.243^{+0.043}_{-0.037}$ & $1.47^{+0.20}_{-0.14}$ & $0.76^{+0.05}_{-0.05}$ &  $100$ \\
   & 2 & 3328.74$\pm$0.44 & 3328.40$\pm$0.44 & $0.187^{+0.022}_{-0.022}$ & $1.01^{+0.14}_{-0.13}$ & $0.55^{+0.03}_{-0.04}$ & $99.6$ \\ \midrule
20 & 0 & 3342.19$\pm$0.10 & 3341.85$\pm$0.10 & $0.353^{+0.040}_{-0.041}$ & $0.91^{+0.15}_{-0.13}$ & $0.71^{+0.04}_{-0.07}$ &  $100$ \\
   & 1 & 3415.97$\pm$0.12 & 3415.62$\pm$0.12 & $0.530^{+0.060}_{-0.062}$ & $1.18^{+0.09}_{-0.15}$ & $0.99^{+0.05}_{-0.05}$ &  $100$ \\
   & 2 & 3485.02$\pm$0.62 & 3484.67$\pm$0.62 & $0.159^{+0.027}_{-0.018}$ & $1.42^{+0.17}_{-0.22}$ & $0.60^{+0.02}_{-0.03}$ & $99.9$ \\ \midrule
21 & 0 & 3497.66$\pm$0.18 & 3497.30$\pm$0.18 & $0.301^{+0.050}_{-0.033}$ & $1.46^{+0.19}_{-0.23}$ & $0.83^{+0.04}_{-0.04}$ &  $100$ \\
   & 1 & 3571.11$\pm$0.11 & 3570.75$\pm$0.12 & $0.451^{+0.074}_{-0.050}$ & $1.71^{+0.13}_{-0.21}$ & $1.10^{+0.04}_{-0.04}$ &  $100$ \\
   & 2 & 3641.19$\pm$0.36 & 3640.82$\pm$0.36 & $0.178^{+0.027}_{-0.018}$ & $1.92^{+0.16}_{-0.18}$ & $0.73^{+0.04}_{-0.03}$ &  $100$ \\ \midrule
22 & 0 & 3652.89$\pm$0.17 & 3652.52$\pm$0.17 & $0.336^{+0.050}_{-0.033}$ & $1.95^{+0.17}_{-0.18}$ & $1.01^{+0.05}_{-0.05}$ &  $100$ \\
   & 1 & 3726.60$\pm$0.13 & 3726.22$\pm$0.13 & $0.503^{+0.075}_{-0.049}$ & $1.83^{+0.22}_{-0.23}$ & $1.20^{+0.05}_{-0.05}$ &  $100$ \\
   & 2 & 3796.59$\pm$0.32 & 3796.20$\pm$0.32 & $0.113^{+0.017}_{-0.013}$ & $1.70^{+0.37}_{-0.31}$ & $0.55^{+0.04}_{-0.04}$ &  $100$ \\ \midrule
23 & 0 & 3808.17$\pm$0.23 & 3807.79$\pm$0.23 & $0.212^{+0.032}_{-0.024}$ & $1.67^{+0.40}_{-0.32}$ & $0.75^{+0.06}_{-0.06}$ &  $100$ \\
   & 1 & 3882.48$\pm$0.26 & 3882.08$\pm$0.26 & $0.319^{+0.048}_{-0.035}$ & $2.80^{+0.36}_{-0.34}$ & $1.19^{+0.05}_{-0.07}$ &  $100$ \\
   & 2 & 3952.93$\pm$0.74 & 3952.53$\pm$0.74 & $0.046^{+0.009}_{-0.010}$ & $3.84^{+0.64}_{-0.61}$ & $0.53^{+0.05}_{-0.08}$ & $99.7$ \\ \midrule
24 & 0 & 3964.72$\pm$0.65 & 3964.32$\pm$0.65 & $0.087^{+0.017}_{-0.019}$ & $4.01^{+0.70}_{-0.66}$ & $0.75^{+0.07}_{-0.11}$ &  $100$ \\
   & 1 & 4038.82$\pm$0.53 & 4038.41$\pm$0.53 & $0.130^{+0.026}_{-0.028}$ & $3.89^{+0.57}_{-0.70}$ & $0.88^{+0.11}_{-0.11}$ &  $100$ \\
   & 2 & 4108.98$\pm$1.17 & 4108.56$\pm$1.17 & $0.033^{+0.011}_{-0.006}$ & $3.72^{+1.15}_{-1.22}$ & $0.43^{+0.05}_{-0.04}$ & $89.3$ \\ \midrule
25 & 0 & 4120.30$\pm$0.62 & 4120.87$\pm$0.62 & $0.063^{+0.021}_{-0.012}$ & $3.66^{+1.31}_{-1.29}$ & $0.59^{+0.07}_{-0.06}$ &  $100$ \\
   & 1 & 4195.00$\pm$0.52 & 4194.57$\pm$0.52 & $0.094^{+0.031}_{-0.017}$ & $3.77^{+0.79}_{-0.44}$ & $0.75^{+0.10}_{-0.07}$ &  $100$ \\
   & 2 & 4264.85$\pm$2.49 & 4264.42$\pm$2.49 & $0.024^{+0.003}_{-0.005}$ & $4.09^{+0.72}_{-0.66}$ & $0.38^{+0.03}_{-0.03}$ & $61.3$ \\ \midrule
26 & 0 & 4276.86$\pm$1.02 & 4276.42$\pm$1.02 & $0.045^{+0.006}_{-0.010}$ & $4.16^{+0.75}_{-0.85}$ & $0.53^{+0.04}_{-0.05}$ & $85.3$ \\
   & 1 & 4351.43$\pm$0.66 & 4340.99$\pm$0.66 & $0.067^{+0.009}_{-0.015}$ & $4.44^{+1.53}_{-1.85}$ & $0.67^{+0.08}_{-0.13}$ &  $100$ \\
\bottomrule
\end{tabular}
\end{table*}

\newpage
\phantom{Text}
\newpage
\phantom{Text}
\newpage
\phantom{Text}
\newpage

\appendix
\section{``Best Fitter'' methodology and priors} \label{app:A}
The fitting procedure had several steps. First, a global fit over the frequency range $\Delta=[100, 6000]$ $\mu$Hz was performed that describes the mode envelopes by two Gaussians (one for each star). The noise background was described by two Harvey-like profiles \citep{Harvey1985} with the exponent initially fixed at 2, plus a white noise component,
\begin{equation}
	N(\nu)= \frac{H_0}{1 + (\tau_0 \nu)^{p_{0}}} + \frac{H_1}{1 + (\tau_1 \nu)^{p_{1}}} + N_0.
\end{equation}

Secondly, following the semi-automatic approach from \cite{Benomar2012MNRAS} we estimated the linewidths\footnote{Mode widths are estimated using Fig. 4b from \cite{White2012}, which shows the empirical relation between the width and the asymptotic frequency phase, $\epsilon$.} and heights. Input frequencies were obtained by rescaling solar frequencies, as explained in Sect.~\ref{sec:mode_param}. We then performed a global fit of all individual modes that are significantly above the noise level. For the purposes of efficiency, this was done only over the frequency range $\Delta=[1500, 4600]$ $\mu$Hz. Note that in this range, the lower-frequency Harvey-like profile does not contribute significantly to the noise background, so it was discarded in further steps.

Priors for all parameters needed to be adequately defined. Two families of priors can be distinguished: (1) generic priors suitable for most solar-like oscillators and (2) priors set after visual inspection of the power spectrum. For convenience we first recall the definition of the indicator function for $x\in \mathbb{R}$, the extend real line, and $a,b \in \mathbb{R}$ and such that $a < b$,
\begin{equation}
\mathds{1}_{[a,b]}(x) = \begin{cases}
1  & \text{if $x \in [a,b]$}\\
0  & \text{otherwise}.
\end{cases}
\end{equation}

\subsection{Generic priors}
Generic priors are those which are not specifically tuned for the HD 176465 binary system, but are suitable for most cool main-sequence solar-like oscillators. 
Table \ref{tab:priors:generic} summarizes the nature of the priors. In this category, two kinds of priors were used: the Jeffrey-truncated prior and the uniform prior \citep{Jeffreys1961}. The Jeffrey-truncated\footnote{Note that this prior is truncated because its integral would be otherwise undefined.} prior,
\begin{equation} \label{eq:Jeffrey1}
	J(x ; a,b) = \frac{\ln(1 + b/a)}{x + a} \mathds{1}_{[a,b]}(x),
\end{equation}
can be appropriate in the case of a scale parameter (i.e. a parameter such as $f(x) = ax$ with $a$, a scale factor), which is the case for the heights and widths of the modes. Note that this prior is non-informative and uniform in the logarithmic domain so that it is invariant under parameter transformation in a logarithmic scale.

The uniform prior defined as,
\begin{equation} \label{eq:Jeffrey2}
	U(x ; a,b) = \frac{1}{b-a} \mathds{1}_{[a,b]}(x),
\end{equation}
is recommended by Jeffreys for location parameters (i.e. parameters such as $f(x) = x + a)$, which is the case, for example, for the mode frequencies or the rotational splitting.

\begin{table}
\centering
\caption{Generic priors and noise priors used during the fit of the mode parameters of the HD 176465 binary system.}
\label{tab:priors:generic}
\begin{tabular}{ccc}
\toprule
Parameter type & Prior  & Unit\\
\midrule
Mode height $H$              & J(1, 250) & ppm$^2/\mu$Hz \\ 
Mode width  $\Gamma$              & J(0.5, 40) & $\mu$Hz \\ 
Mean rotational splitting $\nu_s$     & U(0, 3.5)  & $\mu$Hz \\
Stellar inclination $i$     & U(0, 90)  & degree \\ 
\midrule
Noise height $H_1$   & G(0.417, 0,036)  & ppm$^2/\mu$Hz \\
Noise timescale $\tau_1$ & G(999, 214) & second \\
Noise power $p_1$ & G(2.0, 0.1)  & no unit \\
White noise $N_0$   & G(0.2307, 0.0045) &    ppm$^2/\mu$Hz \\
\bottomrule
\end{tabular}
\end{table}

\subsection{Priors specific to the HD 176465 system}
The parameters of the noise background and oscillation frequencies require specific priors for HD 176465~A~and~B.

Regarding the noise background, we used Gaussian priors defined as $G(x; x_c, \sigma)=\mathrm{e}^{-0.5 (x-x_c)^2/\sigma^2}/\sqrt{2 \pi} \sigma$ for all of the parameters of the noise function, $H_1$, $\tau_1$, $p_1$ and $N_0$.
The characteristic values for $x_c$ and $\sigma$ are given in Table~\ref{tab:priors:generic}.

The priors for the oscillation frequencies must limit the range of acceptable values in order to achieve the correct mode identification. This is especially critical in the case of HD\,176465, as the two components of the binary system oscillate at similar frequencies so that individual modes of the stars may overlap.

One could use uniform priors to define the acceptable range for a given pulsation frequency. However, a trial and error approach over large set of solar-like oscillators showed that choosing uniform priors with Gaussian edges for frequencies defined as, 
\begin{multline}
	F(x ; a,b,\sigma) = C\exp\left[ -\frac{1}{2}\frac{(x-a)^2}{\sigma^2}\right] \mathds{1}_{]-\infty,a[}(x) + C \,\mathds{1}_{[a,b]}(x) \\+ C\exp\left[ -\frac{1}{2}\frac{(x-b)^2}{\sigma^2}\right] \mathds{1}_{]b,+\infty[}(x)
\end{multline}
 enhanced the stability during the training phase of the MCMC process when dealing with a large parameter space \citep[see][for more details about the training phase]{Benomar09}. Here, $C=(b - a + 0.5 \sqrt{2 \pi} \sigma)^{-1}$ is the normalization constant and $\sigma=0.2$ $\mu$Hz. Values of the boundaries $a$ and $b$ for each mode are given in table \ref{tab:priors_nu: Luke} for HD\,176465\,A and \ref{tab:priors_nu: Leia} for HD\,176465\,B.

The fit of HD\,176465 was complicated by the overlapping frequencies of the stars. This is evident from the \'echelle diagrams in Fig. \ref{fig:ech}. When this occurs, it is not possible to unambiguously separate the power due to one star from the other unless further priors are defined. As described by Equation (\ref{eqn:asympt}), $p$ modes of each star are approximately regularly spaced, so that the second derivative in radial order n of that equation is $\frac{\partial^2 \nu(l,n)}{\partial n^2} \approx 0$. Thus one may limit local strong deviations from the regular pattern by imposing a Gaussian prior so that $\frac{\partial^2 \nu(l,n)}{\partial n^2} = G(0, \sigma)$. Numerically, the second derivative was performed using a forward/backward difference on the edges and an entered difference otherwise. A value of $\sigma=2$ $\mu$Hz/n$^2$ was found to be suitable to de-correlate the overlapping modes.

\begin{table}
\centering
\caption{Frequency priors for HD\,176465\,A, along with the fitted frequency. The prior is uniform in the range $[x_\mathrm{min}, x_\mathrm{max}]$ but has Gaussian edges with standard deviation $\sigma=0.2$ $\mu$Hz.}
\label{tab:priors_nu: Luke}
\begin{tabular}{ccccc}
\toprule
$n$ & $\ell$ & Output Frequency ($\mu$Hz) & $[a, b]$ \\ \midrule
12 & 2 & 2121.39$\pm$1.67 &   [2116.99, 2134.32] \\  \midrule
13 & 0 & 2132.67$\pm$0.61 &  [2127.84, 2135.17] \\
   & 1 & 2202.44$\pm$0.39 &  [2197.82  2204.35] \\
  & 2 &  2268.74$\pm$1.17 & [2262.39, 2281.03] \\  \midrule
14 & 0 & 2280.58$\pm$0.90 &  [2269.85,  2285.97] \\
   & 1 & 2348.98$\pm$0.24 &  [2339.54, 2358.89] \\  
   & 2 & 2415.63$\pm$1.97 &  [2410.13, 2430.27] \\  \midrule
15 & 0 & 2428.06$\pm$0.38 &  [2423.23,  2440.56]\\
  & 1 & 2496.73$\pm$0.33 &  [2491.47, 2503.45]\\ 
  & 2 & 2561.06$\pm$1.65 &  [2555.72, 2577.18] \\  \midrule
16 & 0 & 2575.83$\pm$0.83 & [2565.60, 2587.66] \\ 
   & 1 & 2642.66$\pm$0.10 &  [2640.57, 2646.71] \\ 
   & 2 & 2706.15$\pm$0.74 &  [2698.51, 2717.15] \\  \midrule
17 & 0 & 2720.12$\pm$0.28 & [2713.56, 2723.17] \\
   & 1 & 2788.08$\pm$0.36 & [2783.74, 2793.14] \\ 
   & 2 & 2852.75$\pm$0.38 & [2846.03, 2858.21]\\  \midrule
18 & 0 & 2865.49$\pm$0.09 & [2863.66, 2867.32] \\
   & 1 & 2934.23$\pm$0.12 & [2931.46, 2937.10] \\ 
   & 2 & 2999.34$\pm$0.22 & [2991.83, 3005.24] \\   \midrule 
19 & 0 & 3012.10$\pm$0.11 & [3006.55, 3014.44] \\
   & 1 & 3081.19$\pm$0.10 &  [3077.39, 3086.70]\\ 
   & 2 & 3146.38$\pm$0.21 &  [3139.26, 3153.16] \\  \midrule
20 & 0 & 3158.74$\pm$0.02 & [3154.07, 3162.85] \\
    & 1 & 3227.97$\pm$0.09 & [3224.83, 3230.38] \\ 
   & 2 & 3293.79$\pm$0.17 &  [3289.20, 3299.27] \\  \midrule
21 & 0 & 3305.62$\pm$0.08 &  [3301.69, 3310.76] \\
    & 1 & 3375.32$\pm$0.10 &  [3369.81, 3380.09] \\ 
   & 2 & 3441.16$\pm$0.36 &  [3433.29, 3448.81] \\  \midrule
22 & 0 & 3452.35$\pm$0.11 &  [3447.70, 3457.17] \\
   & 1 & 3522.07$\pm$0.14 &  [3514.91, 3528.31] \\ 
   & 2 & 3588.47$\pm$0.26 &  [3583.73, 3593.71] \\  \midrule
23 & 0 & 3599.69$\pm$0.13 & [3593.41, 3605.80] \\
   & 1 & 3669.46$\pm$0.16 &  [3664.65, 3675.13] \\ 
   & 2 & 3735.42$\pm$0.63 &  [3731.54, 3743.92] \\  \midrule
24 & 0 & 3746.85$\pm$0.16 & [3742.64, 3752.11]\\
   & 1 & 3817.25$\pm$0.24 &  [3812.14, 3822.83] \\ 
   & 2 & 3884.21$\pm$1.72 &  [3876.05, 3892.18] \\  \midrule
25 & 0 & 3895.17$\pm$0.18 &  [3890.94, 3900.05] \\
   & 1 & 3965.06$\pm$0.40 &  [3959.44, 3971.12] \\ 
   & 2 & 4033.13$\pm$1.44 &  [4022.67,  4045.10] \\  \midrule
26 & 0 & 4043.26$\pm$0.02 &  [4032.42, 4048.46] \\
   & 1 & 4114.50$\pm$0.50 &  [4109.93, 4121.31]\\  
   & 2 & 4181.86$\pm$0.89 &  [4173.92, 4190.65] \\  \midrule
27 & 0 & 4192.29$\pm$0.46 &  [4187.77, 4202.42] \\
   & 1 & 4263.57$\pm$0.17 &  [4252.20, 4273.98] \\ 
   & 2 & 4329.12$\pm$2.46 &  [4322.77, 4343.90] \\  \midrule
28 & 0 & 4341.82$\pm$0.91 &  [4334.49, 4344.79] \\
  & 1 & 4411.38$\pm$0.27 &  [4394.87, 4417.14]\\
\bottomrule
\end{tabular}
\end{table}

\begin{table}
\centering
\caption{Frequency priors for HD\,176465\,B, along with the fitted frequency. The prior is uniform in the range $[x_\mathrm{min}, x_\mathrm{max}]$ but has Gaussian edges with standard deviation $\sigma=0.2$ $\mu$Hz.}
\label{tab:priors_nu: Leia}
\begin{tabular}{ccccc}
\toprule
$n$ & $\ell$ & Output Frequency ($\mu$Hz) & $[a, b]$ \\ \midrule
15 & 2 & 2709.26$\pm$2.31 & [2706.26, 2712.74] \\  \midrule
16 & 0 & 2725.97$\pm$1.56 & [2721.93,  2729.03] \\
   & 1 & 2797.21$\pm$0.43 &  [2793.96, 2799.60] \\
   & 2 & 2864.06$\pm$1.63 & [2858.83, 2874.60] \\  \midrule
17 & 0 & 2879.67$\pm$0.47 & [2874.39,  2882.85]\\
   & 1 & 2951.76$\pm$0.21 & [2947.15  2955.30]  \\
   & 2 & 3018.83$\pm$1.24 & [3014.32, 3027.06]  \\  \midrule
18 & 0 & 3032.77$\pm$0.42 & [3027.58, 3037.39] \\
   & 1 & 3105.54$\pm$0.21 & [3099.83, 3113.54]\\
   & 2 & 3173.61$\pm$0.71 &  [3166.46, 3181.40]\\  \midrule
19 & 0 & 3187.74$\pm$0.26 &  [3182.02, 3192.57] \\
   & 1 & 3260.58$\pm$0.15 & [3255.83, 3263.87] \\
   & 2 & 3328.74$\pm$0.44 & [3325.40, 3337.20] \\ \midrule
20 & 0 & 3342.19$\pm$0.10 & [3338.90, 3349.95] \\
   & 1 & 3415.97$\pm$0.12 & [3408.26, 3422.71]\\
   & 2 & 3485.02$\pm$0.62 &  [3479.63, 3494.25]  \\  \midrule
21 & 0 & 3497.66$\pm$0.18 & [3494.04, 3500.83] \\
   & 1 & 3571.11$\pm$0.11 & [3567.12, 3575.47]\\
  & 2 & 3641.19$\pm$0.36 &  [3634.79, 3648.60] \\ \midrule
22 & 0 & 3652.89$\pm$0.17 & [3648.49, 3656.32] \\
   & 1 & 3726.60$\pm$0.13 & [3722.61, 3731.03]\\
   & 2 & 3796.59$\pm$0.32 & [3789.15, 3803.87] \\ \midrule
23 & 0 & 3808.17$\pm$0.23 &  [3804.39, 3811.49] \\
   & 1 & 3882.48$\pm$0.26 & [3877.57, 3886.14]\\
   & 2 & 3952.93$\pm$0.74 & [3947.14, 3959.36] \\ \midrule
24 & 0 & 3964.72$\pm$0.65 & [3959.25, 3970.32] \\
   & 1 & 4038.82$\pm$0.53 & [4034.73, 4044.13] \\
   & 2 & 4108.98$\pm$1.17 & [4105.34, 4118.61] \\ \midrule
25 & 0 & 4120.30$\pm$0.62 & [4116.52, 4128.74]\\
   & 1 & 4195.00$\pm$0.52 &  [4192.20, 4199.83]\\
   & 2 & 4264.85$\pm$2.49 & [4262.82,  4273.99] \\ \midrule
26 & 0 & 4276.86$\pm$1.02 & [4270.65, 4282.76] \\
   & 1 & 4351.43$\pm$0.66 &  [4347.48, 4356.78] \\
\bottomrule
\end{tabular}
\end{table}

\subsection{Mode significance}
The mode significance was determined by locally fitting each mode of frequency $\nu(n,l)$ over $\Delta = [\nu(n,l) - \Delta\nu/3, \nu(n,l) + \Delta\nu/3]$, where $\Delta\nu$ is the large separation. Two fits were required to make a model comparison possible. The first fit uses a model $M_{mode}$ that included the mode for which the probability was calculated, along with a noise model. The second fit included only the noise model $M_{noise}$. Using Equation (\ref{eq:posterior_model}), the Bayes factor was calculated assuming the {\it a priori} equiprobability of the two models, $\pi(M_{noise})=\pi(M_{mode}) = 0.5$, so that
\begin{equation}
	P(M_{mode}|y)=\frac{1}{1 + \pi(y| noise)/\pi(y|mode)}.
\end{equation}

More details about the technical implementation to calculate $\pi(y| noise)$ and $\pi(y| mode)$ can be found in \cite{Benomar09b}.
Note that within $\Delta$, modes other than the one for which we wished to calculate the probability could be present. These were included as a part of the noise background, with parameters fixed to their median values obtained during the global fit. Note also that the two fits were made with the same priors as for the global fit. 

\clearpage

\end{document}